\documentclass[10pt]{iopart}

\newcommand{\HII}{H\,{\scriptsize II}}

\newcommand{\CII}{[C\,{\scriptsize II}]}
\newcommand{\twCII}{[$^{12}$C\,{\scriptsize II}]}
\newcommand{\tCII}{[$^{13}$C\,{\scriptsize II}]}

\newcommand{\OI}{[O\,{\scriptsize I}]}

\newcommand{\CI}{[C\,{\scriptsize I}]}

\newcommand{\micron}{$\mu$m}

\usepackage{hyperref}
\usepackage{graphicx}
\usepackage{txfonts}
\usepackage{natbib}
\usepackage[multiple]{footmisc}

\begin{document}

\title[The SOFIA legacy project FEEDBACK]{FEEDBACK: a SOFIA Legacy Program to Study Stellar Feedback in Regions of Massive Star Formation}
\vspace{0.5cm}

\noindent \textsc{N. Schneider\footnote{I. Physik. Institut, University of Cologne, Z\"ulpicher Str. 77, D-50937 Cologne, Germany},
  R. Simon$^1$,
  C. Guevara$^1$,
  C. Buchbender$^1$, 
  R.D. Higgins$^1$,
  Y. Okada$^1$,
  J. Stutzki$^1$,
  R. G\"usten\footnote{Max-Planck Institut f\"ur Radioastronomie, Auf dem H\"ugel 69, D-53121 Bonn, Germany},
  L.D. Anderson\footnote{Department of Physics and Astronomy, West Virginia University, Morgantown WV 26506, USA},
  J. Bally\footnote{Center for Astrophysics and Space Astronomy, University of Colorado, Colorado 80309, USA},
  H. Beuther\footnote{Max Planck Institute for Astronomy, K\"onigstuhl 17, D-69117 Heidelberg, Germany}, 
  L. Bonne\footnote{LAB, University of Bordeaux, CNRS, B18N, F-33615 Pessac, France},
  S. Bontemps$^6$,
  E. Chambers\footnote{USRA/SOFIA, NASA Ames Research Center, Moffett Field, CA 94035-0001, USA},
  T. Csengeri$^6$,
  U.U. Graf$^1$, 
  A. Gusdorf\footnote{LPENS, Universit\'e PSL, CNRS, Sorbonne Universit\'e, Universit\'e de Paris, Paris, France},
  K. Jacobs$^1$, 
  M. Justen$^1$, 
  S. Kabanovic$^1$,
  R. Karim\footnote{Department of Astronomy, University of Maryland, College Park, MD 20742, USA},
  M. Luisi$^3$, 
  K. Menten$^2$,
  M. Mertens$^1$, 
  B. Mookerjea\footnote{Tata Institute of Fundamental Research, Homi Bhabha Road, Mumbai 400005, India}, 
  V. Ossenkopf-Okada$^1$,
  C. Pabst\footnote{Leiden Observatory, Leiden University, PO Box 9513, 2300 RA Leiden, The Netherlands}, 
  M.W. Pound$^9$,
  H. Richter\footnote{Institute of Optical Sensor Systems, DLR, Rutherfordstr. 2, D-12489 Berlin, Germany},  
  N. Reyes$^{2}$, 
  O. Ricken$^{2}$, 
  M. R\"ollig$^1$,
  D. Russeil\footnote{Aix Marseille Universit\'e, CNRS, CNES, LAM, Marseille, France},
 \'A. S\'anchez-Monge$^1$,
  G. Sandell\footnote{Institute for Astronomy, University of Hawaii, 640 N. Aohoku Place, Hilo, HI 96720, USA},
  M. Tiwari$^{9}$, 
  H. Wiesemeyer$^2$, 
  M. Wolfire$^{9}$,
  F. Wyrowski$^2$, 
  A. Zavagno$^{13}$,
  A.G.G.M. Tielens$^{9,11}$} 

\ead{nschneid@ph1.uni-koeln.de}


\begin{abstract}
FEEDBACK is a SOFIA (Stratospheric Observatory for Infrared Astronomy)
legacy program dedicated to study the interaction of
massive stars with their environment.  It performs a survey of 11
galactic high mass star forming regions in the 158 $\mu$m (1.9 THz)
line of \CII\ and the 63 $\mu$m (4.7 THz) line of \OI. We employ the
14 pixel Low Frequency Array (LFA) and 7 pixel High Frequency Array
(HFA) upGREAT heterodyne instrument to spectrally resolve (0.24 MHz)
these far-infrared fine structure lines.
With a total observing time of 96h, we will cover
$\sim$6700 arcmin$^2$ at 14.1$''$ angular resolution for the
\CII\ line and 6.3$''$ for the \OI\ line. The observations started in
spring 2019 (Cycle 7).  Our aim is to understand
the dynamics in regions dominated by different feedback processes from
massive stars such as stellar winds, thermal expansion, and radiation
pressure, and to quantify the mechanical energy injection and
radiative heating efficiency. This is an important science topic
because feedback of massive stars on their environment regulates the
physical conditions and sets the emission characteristics in the
interstellar medium (ISM), influences the star formation activity
through molecular cloud dissolution and compression processes,
and drives the evolution of the ISM in
galaxies.  The \CII\ line provides the kinematics of the gas 
and is one of the dominant cooling lines of gas for low to moderate densities and UV
fields. The \OI\ line traces warm and high-density gas, excited in
photodissociations regions with a strong UV field or by shocks.  The
source sample spans a broad range in stellar characteristics from
single OB stars, to small groups of O stars, to rich young stellar
clusters, to ministarburst complexes. It contains well-known targets
such as Aquila, the Cygnus X region, M16, M17, NGC7538, NGC6334, Vela,
and W43 as well as a selection of \HII\ region bubbles, namely RCW49,
RCW79, and RCW120.  These \CII\ maps, together with the less explored
\OI\ 63 \micron\ line, provide an outstanding database for the
community.  They will be made publically available and will
trigger further studies and follow-up observations.
\end{abstract}

%
\vspace{2pc}
\ioptwocol

\section{Introduction} \label{introduction}

The interaction of massive stars with their environments regulates the
evolution of galaxies. Mechanical and radiative energy input by
massive stars stir up and heat the gas and control cloud and
intercloud phases of the interstellar medium (ISM).  Stellar feedback
also governs the star formation efficiency of molecular clouds
\citep{elmegreen2011,hopkins2014}. On the one hand, stellar feedback
can lead to a shredding of the nascent molecular cloud within a few
cloud freefall times thereby halting star formation
\citep{matzner2002,geen2016,kim2018}. The efficiency of this process
is somewhat controversial as some models \citep{dale2014}, combining
effects of photoionization and momentum-driven winds, conclude that
the effectiveness of stellar feedback to disperse a cloud is small. A
recent observational study of \citet{watkins2019} argues similarly,
proposing that feedback does not influence dense gas and that the
cloud morphology and average density at the time when the first
O-stars form is decisive for the star-formation efficiency.  On the
other hand, it has been proposed that massive stars can also provide
positive feedback to star formation as gravity can more easily
overwhelm cloud-supporting forces in swept-up compressed shells
\citep{elmegreen1977}. A popular observational example for such a
triggering of star formation is RCW120
\citep{deharveng2010,zavagno2010}. But again, this scenario was
challenged by \citet{walch2015} who propose a hybrid form of
triggering, which combines elements of collect \&\ collapse
\citep{elmegreen1977} and radiation driven implosion
\citep{bertoldi1989,lefloch1994} models.

Separate from the issue of how stellar feedback impacts
star-formation, it is evident that stars control the radiative energy
budget of the ISM and its emission characteristics. Extreme
ultra-violet (EUV) photons from massive stars with energies $>$13.6 eV
ionize hydrogen, creating \HII\ regions that cool through H
recombination lines and forbidden, collisionally excited transitions
of trace elements. Less energetic photons photodissociate molecules
and photoionize species such as carbon (C) and sulfur (S). These
far-UV (FUV) photons heat the gas through photoelectrons from large
Polycyclic Aromatic Hydrocarbon (PAH) molecules and very small grains, and
by collisional de-excitation of vibrationally excited H$_2$ molecules, 
while it cools predominantly through far-infrared fine-structure lines
of ionized carbon (C$^+$) and atomic oxygen (O).  The result is a 
Photo Dissociation Region (PDR): a layer of warm, atomic and
molecular gas that separates the ionized gas from the surrounding
molecular cloud material \citep{hollenbach1999}. The physical
properties of PDRs are largely controlled by the coupling of the FUV
photons to the energy budget of the gas. Detailed models have been
developed for this coupling \citep{bakes1994} but as the properties of
large molecules and very small grains are not well known, these models
are uncertain. Observational characterization of the heating
efficiency of FUV photons has hitherto only been possible for a
handful of sources \citep{okada2013,pabst2017}. This uncertainty in
the heating of the gas enters directly into PDR models 
\citep{kaufman2006,roellig2006,roellig2007,lepetit2006,pound2008} that
are widely used to analyze observations and, as a result, the derived
physical properties carry a large systematic uncertainty.

Another source of ionization and heating must be considered when
massive stars are present. These are X-rays, produced when the fast
stellar wind of OB stars shocks the surrounding medium
\citep{weaver1977}.  Molecular gas exposed to X-rays has a different
chemical structure and thermal balance than PDRs. These X-ray
dominated regions (XDRs) span a temperature range of $\sim$200~K to
10$^4$~K and produce large column densities of warm gas
\citep{maloney1996}.  In general, the mechanical luminosity of the
stellar wind is only a small ($\sim$10$^{–3}$) fraction of the
radiative luminosity and the X-ray luminosity is only a small
($\sim$10$^{–4}$) fraction of the mechanical luminosity (most of the
cooling goes through adiabatic expansion). Hence, X-rays will probably
not affect significantly the typical PDR cooling lines (\CII, \OI, low
to mid-J CO).  Modeling emission lines arising in PDRs and XDRs
\citep{meijerink2006,spaans2008} predict that high-J CO lines are good
XDR tracers. However, high density, high temperature PDR models also
succeed in explaining observed CO line fluxes \citep{stock2015}.  In
order to separate the PDR contribution to XDR diagnostic tracers
requires a deep understanding of the physics of FUV heated gas and its
cooling response.

Cosmic Rays (CRs) represent yet another channel of feedback exerted
from stars on the ISM of galaxies. The bulk of the mid-energy
($\sim$1~GeV) CRs is thought to be originating from supernova (SN)
explosions and supernova remnants \citep{bykov2018}, but the idea that
large (typically $>$10~pc) superbubbles could contribute significantly
to the Galactic CR budget has recently emerged \citep{tatischeff2018}.
Though most of CRs quickly diffuse away, they may still play a role in
the FUV shielded \HII\ region/molecular cloud interface
\citep{padovani2019,meng2019}.  We will thus make effort to consider
and study the impact of CRs on the immediate environment of massive
stars.

Through their winds and explosions, massive stars impact the ISM
dynamically on all scales. Turbulence is driven on the sub-parsec
scale by radiation pressure and jets and outflows, but can also be
generated Galaxy-wide by supernovae explosions. This injection of
mechanical energy into the ISM is the origin of the Hot Intercloud
Medium (HIM) and a source of turbulent pressure that supports the gas
disk and the clouds therein against self-gravity and gravitational
collapse in the host galaxy's potential \citep{mckee1977}.
%
At intermediate scales (up to a few tens of parsec), radiation
pressure and winds sweep up gas into 'bubbles', i.e., ring- or
shell-like structures. Surveys such as the mid-IR Galactic Legacy
Infrared Mid-Plane Survey Extraordinaire (GLIMPSE) have revealed that
these bubbles are ubiquitous morphological features in the ISM of the
Milky Way \citep{churchwell2006} and enclose \HII\ regions
\citep{deharveng2009}.  They may be caused by the thermal expansion of
\HII\ regions driven by the overpressure of ionized gas, by the
mechanical action of stellar winds from massive stars creating X-ray
emitting hot gas, and by the effects of radiation pressure on
surrounding dust and gas
\citep{stroemgren1939,spitzer1968,weaver1977,draine2011,haid2016}.
However, it is not settled whether these bubbles are indeed 3D
structures.  In the plane of the sky, they resemble ring-like
structures but this geometry might largely reflect projection effects
\citep{beaumont2010,kirsanova2019}. Recent \CII\ observations of the
Orion molecular cloud, on the other hand, clearly reveal 3D expanding
shell structures associated with the areas of active star formation,
M42, M43, and NGC 1977 \citep{pabst2019,pabst2020}.

Not all \HII\ regions have a bubble morphology. In particular massive
star-forming giant molecular cloud complexes such as the Cygnus X
region or W43 are a mixture of individual bubble-like \HII\ regions
and a more inhomogeneous distribution of ionized gas in which the cool
molecular gas is pervaded by a web of filaments and larger clouds.
Under the influence of ionizing radiation, a rich diversity of
structures such as pillars, bright-rimmed clouds and globules are created at
the \HII\ region/molecular cloud interface.  These features have long been 
observed at various wavelengths 
\citep{herbig1974,schneps1980,hester1996,white1997,schneider2016}, but
it is mostly unexplored how radiation and stellar wind impact
filaments and massive cloud ridges, and how that influences subsequent
filament/cloud evolution and star formation. So far, it is known, 
thanks to far-infrared dust imaging surveys by {\sl Herschel}, that
filaments play an important role in the molecular cloud- and
star-formation process
\citep{andre2010,schneider2012a,hill2012b,hennemann2012}.  Simulations
performed by \citet{inutsuka2015} show that molecular clouds can form
due to multiple compressions by overlapping dense shells driven by
expanding bubbles. Filaments are then the first step of this formation
process, i.e., the compressed shell, as was proposed in a recent study
of the RCW120 bubble \citep{zavagno2020}. Summarizing, detailed
studies of both small, bubble-like \HII\ regions and large cloud complexes
are required to probe the underlying physical processes, the structure
and physical properties of the ambient ISM into which they are
expanding, the hydrodynamical response of gas in the ISM to stellar
action, and the radiative coupling of gas and dust to the intense
photon fields of massive stars.

The efficient mapping capabilities of ground based sub-mm wavelength 
radio-telescopes, the upGREAT array receiver on SOFIA, and (F)IR imaging
from {\sl Spitzer}, {\sl WISE}, and {\sl Herschel} has now enabled in depth
studies of the coupling between molecular clouds and OB stars.
In particular the \CII\ $^2$P$_{3/2}$-$^2$P$_{1/2}$ fine-structure line at
157.74 \micron\ or 1.90054 THz ($E/k_b$=91.2~K) offers a unique probe of the radiative
and kinetic interaction of massive stars with their environment. The
C$^+$ ion is the dominant form of carbon in atomic hydrogen and
CO-dark molecular gas layers because it has a low ionization potential
of 11.3~eV, slightly below the ionization potential of hydrogen which
is 13.6~eV.  The \CII\ line is the dominant cooling line of PDR
surfaces of FUV illuminated molecular clouds where stellar photons
dissociate molecular gas and heat it to temperatures of
T$\sim$200~K.
The line is easy to excite thermally by collisions with
electrons, atomic hydrogen and molecular hydrogen. The critical
density, defined by the collisional de-excitation rate being equal to
the effective spontaneous decay rate, depends on the temperature.  It
is 9~cm$^{-3}$, 3$\times$10$^3$~cm$^{-3}$, and 6.1$\times$10$^3$~cm$^{-3}$
for collisions with e$^-$, H and H$_2$, respectively, for gas temperatures
$\lesssim$100 K \citep{goldsmith2012}. \\ 
The \OI\ line at 63.18 \micron\ or 4.74478 THz ($E/k_b$=228~K) is the most important
cooling line for warmer T$>$200~K and denser gas with a critical
density of 5$\times$10$^5$~cm$^{-3}$ for collisions with H$_2$
\citep{roellig2006}. The ratio of \CII\ to \OI\ line fluxes depends on the
temperature and density of the gas. \\ 
In massive star-forming regions, 
energy is also injected in the form of shocks (expanding ionization
fronts, stellar winds) so that in particular the 
\OI\ observations also need to be interpreted in the context of
irradiated shock models, where the effects of FUV photons on the
physical conditions and chemistry of the region are included.
For understanding the dynamics of the gas emitting the \CII\ 158
\micron\ line, it is also important to carefully study the line
profile in order to assess possible optical depth effects.  Recent
studies \citep{graf2012,ossenkopf2015,mookerjea2018,mookerjea2019}
have shown that the \CII\ line is often optically thick and shows
self-aborption effects.  The \twCII\ optical depth can be
estimated when the \tCII\ hyperfine structure (hfs) satellite emission
is detected \citep{ossenkopf2013,okada2020,guevara2020}.

The SOFIA legacy program FEEDBACK is designed to probe the radiative
and mechanical interaction of massive stars with their natal
clouds. The focus is on understanding how massive stars control star
formation, to what extent they disrupt molecular clouds, how
turbulence is injected into the ISM, if and how new star formation can
be triggered, and how their FUV photons couple to atomic and molecular
gas in the surrounding PDRs. With the 14 pixel Low Frequency Array (LFA)
upGREAT heterodyne spectrometer, we efficiently observe over large scales
(several 1000~arcmin$^2$) and at high spatial (14.1$''$) and high
spectral (0.24~MHz) resolution a statistically significant sample of
Galactic star-forming molecular clouds.  In parallel, the 4.7 THz High
Frequency Array (HFA) channel of upGREAT delivers maps of the \OI\ 63
\micron\ line, which provide the community with unique complementary
data in the context of PDR and shock modeling. These latter maps are
undersampled because the focus was on \CII\ mapping, but were observed
in a regular mapping scheme (Sec.~\ref{observations}).

In order to prepare the community for the opportunities offered by the
FEEDBACK project and to allow them to prepare ancillary observations,
we outline the goals of the project (Section \ref{goals}) and describe
the source selection (Section \ref{sources}). The planning of the
observations, performance and data reduction is presented in Section
\ref{observations}. First results from SOFIA cycle 7 observations are
given in Section \ref{results}. In Section \ref{products} we outline
which data products will be delivered to the community and Section
\ref{conclusions} summarizes the paper.

\begin{table*}  
\caption{FEEDBACK sources information.} \label{table:summary1}   
\begin{center}  
\begin{tabular}{lcccccccccc}  
\hline \hline   
           &                     &                      &                      &                       &            &         &       &\\ 
Cloud      & $\alpha_{J2000}$(ON) &  $\delta_{J2000}$(ON) & $\alpha_{J2000}$(OFF) &  $\delta_{J2000}$(OFF) & d          & \#Tiles & Area     \\           
&  ($^{h}$:$^{m}$:$^{s}$) & ($^{\circ}$:$^{\prime}$:$^{\prime\prime}$) & ($^{h}$:$^{m}$:$^{s}$) & ($^{\circ}$:$^{\prime}$:$^{\prime\prime}$) & \small{(kpc)} &  & \small{(arcmin$^2$)} &    \\  
           & (1)                 & (2)                  &  (3)                 & (4)                   & (5)        & (6)     & (7)   & (8) \\   
\hline       
Cygnus X   & 20:38:20.22         &  42:24:18.29         &  20:39:48.34         &  42:57:39.11           & 1.4$^a$   &  18     &  1262 & PMD  \\ 
M16        & 18:18:35.69         & -13:43:30.98         &  18:17:17.76         & -14:01:42.73           & 1.74$^b$  &   12    &   630 & PMD/CHC \\ 
M17        & 18:20:43.16         & -16:06:14.87         &  18:22:01.14         & -16:21:11.35           & 1.98$^c$  &   13    &   841 & PMD/CHC \\     
NGC6334    & 17:20:14.07         & -35:55:05.18         &  17:18:16.38         & -35:52:28.08           & 1.3$^d$   &   16    &   788 & CHC     \\  
NGC7538    & 23:13:46.41         &  61:31:42.01         &  23:10:54.84         &  61:28:48.03           & 2.65$^e$  &    4    &   210 & PMD     \\ 
RCW49      & 10:24:11.57         & -57:46:42.50         &  10:27:17.42         & -57:13:42.69           & 4.21$^f$  &   12    &   788 & CHC     \\  
RCW79      & 13:40:05.86         & -61:42:36.94         &  13:42:08.76         & -61:15:08.40           & 4.2$^g$   &    9    &   473 & CHC     \\  
RCW120     & 17:12:22.82         & -38:26:51.61         &  17:10:41.01         & -37:44:03.60           & 1.68$^b$  &    4    &   210 & CHC     \\  
RCW36      & 08:59:26.81         & -43:44:14.06         &  09:01:31.67         & -43:22:50.00           &  0.95$^h$ &    4    &   210 & CHC     \\  
W40        & 18:31:28.58         & -02:07:35.39         &  18:33:08.90         & -02:21:36.40           &  0.26$^i$ &   12    &   630 & PMD/CHC \\     
W43        & 18:48:01.04         & -01:58:22.27         &  18:49:30.26         & -02:03:14.94           &  5.49$^j$ &   12    &   630 & PMD/CHC \\  
\hline   
\end{tabular}  
\end{center}  
\vskip0.1cm  
\noindent (1,2) Coordinates of the central position of the \CII\ map (see Fig.~1). 
\noindent (3,4) Coordinates of the emission free reference position. 
\noindent (5) Distance in kpc.
\noindent (6) Number of tiles (one tile is 7.2$' \times$7.2$'$) for mapping. 
\noindent (7) Size of mapping area in arcmin$^2$. 
\noindent (8) Visible from Palmdale (PMD) or Christchurch/New Zealand (CHC). \\
$^a$\citet{rygl2012},
$^b$\citet{kuhn2019},
$^c$\citet{wu2019},
$^d$\citet{chibueze2014},
$^e$\citet{moscadelli2009},
$^f$\citet{cantat2018},
$^g$\citet{russeil1998},
$^h$\citet{massi2019},
$^i$\citet{ortiz2017},
$^j$\citet{zhang2014}.
\end{table*}  

\begin{table*}  
\caption{FEEDBACK sources physical properties.} \label{table:summary2}   
\begin{center}  
\begin{tabular}{lccccccccc}  
\hline \hline   
           &      &                       &         &      &      &          &        &    \\ 
Cloud      & $\langle FUV \rangle$& FUV-range & SpT & v$_{lsr}$ & Mass  & Cloud Geometry & \HII\ Region Geometry       \\           
           &  \small{G$_0$}               & \small{G$_0$} & &  \small{(km s$^{-1})$} & \small{(10$^3$ M$_\odot$)} &        \\  
           & (1)  &  (2)                  & (3)           & (4)  & (5)      & (6)    &  (7)  \\   
\hline       
Cygnus X   &  290 & 200-7.6$\times$10$^4$ &$\sim$50 O, 3 WR &  -3  & 200 & \small{ridge/filaments} & \small{bubbles and compact}   \\ 
M16        &  300 & 270-8.7$\times$10$^3$ & 1 O4, $\sim$10 late O   &  25  &  87 & \small{radiation sculpted interfaces} & \small{irregular} \\ 
M17        & 1295 & 300-4.6$\times$10$^4$ & 2 O4, $\sim$10 late O   &  22  & 483 & \small{clumpy}   & \small{irregular}             \\     
NGC6334    &  580 & 500-8.9$\times$10$^4$ & 5 O5-8, 8 B   &   8  & 239 & \small{ridge/filaments} & \small{many bubbles}             \\  
NGC7538    &  904 & 10$^3$-3.9$\times$10$^5$ & 1 O3             & -55  & 130 & \small{clumpy, ring }   & \small{evolved bubble}      \\ 
RCW49      &  555 & 200-5$\times$10$^3$ & 2 WR, 12 early O &   0  & 170$^+$ & \small{fragmented }     & \small{irregular/evolved bubble}                \\  
RCW79      &  140 & 35-5$\times$10$^3$  & 2 O4, $\sim$10 late O   & -50  &  146 & \small{fragmented}      & \small{evolved bubble}   \\  
RCW120     &  375 & 100-1.5$\times$10$^4$ & 1 O8             & -10  &  20  & \small{ring/shell}      & \small{bubble}                \\  
RCW36      &  413 & 300-1.5$\times$10$^4$ & 1 O8, B-cluster  &  5   & 58  & \small{ridge}           & \small{bipolar}                \\  
W40        &  237 & 150-8.2$\times$10$^3$ & 1 O, 2 B         &  5   & 26  & \small{clumps/filaments}& \small{bipolar}                \\     
W43        &  741 & 700-5.9$\times$10$^4$ & OB, WR cluster   & 100  & 6000 & \small{ridge/filaments}& \small{bubbles and compact}    \\  
\hline   
\end{tabular}  
\end{center}  
\vskip0.1cm  
\noindent (1) Spatially averaged FUV field in Habing units, determined from {\sl Herschel} FIR-fluxes at 70 and 160 $\mu$m \citep{schneider2016}.   
\noindent (2) FUV range in Habing units within the \CII\ mapping area, determined from  {\sl Herschel} FIR-fluxes at 70 and 160 $\mu$m.  
\noindent (3) Spectral type(s) and number of dominating star(s).  
\noindent (4) Approximate local-standard-of-rest velocity of the source. 
\noindent (5) Approximate mass of the associated molecular cloud derived from {\sl Herschel} dust column density maps (Schneider et al. 2020, in preparation).$^+$Mass
determined from CO 1$\to$0 data \citep{furukawa2009}. 
\noindent (6) Type/geometry of the molecular cloud region and interface. 
\noindent (7) Type/geometry of the \HII\ region. 
\end{table*} 

\begin{figure*}[ht]
\begin{center}
\includegraphics[angle=0,width=5.1cm,height=7cm,keepaspectratio]{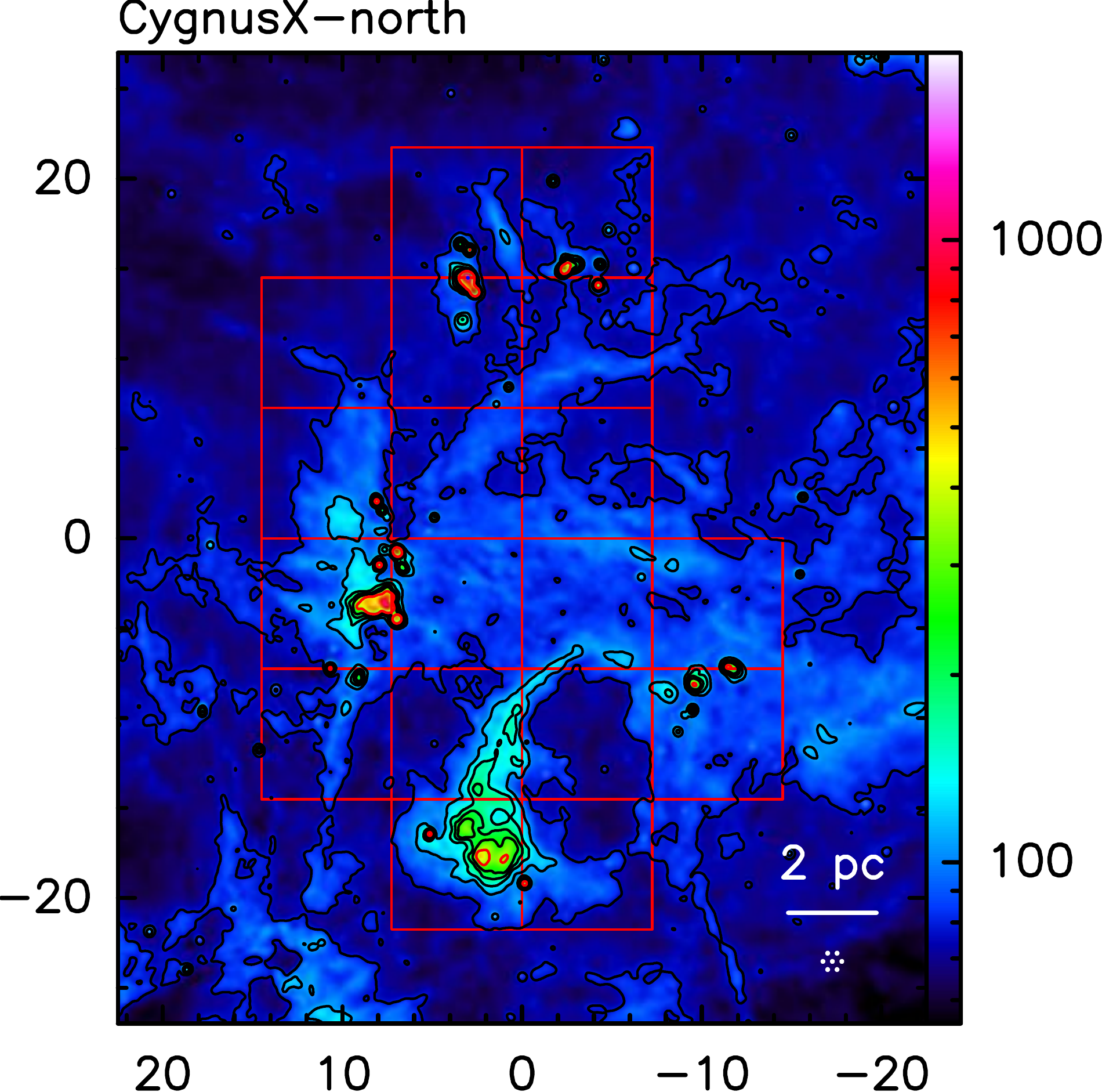} 
\includegraphics[angle=0,width=5.1cm,height=7cm,keepaspectratio]{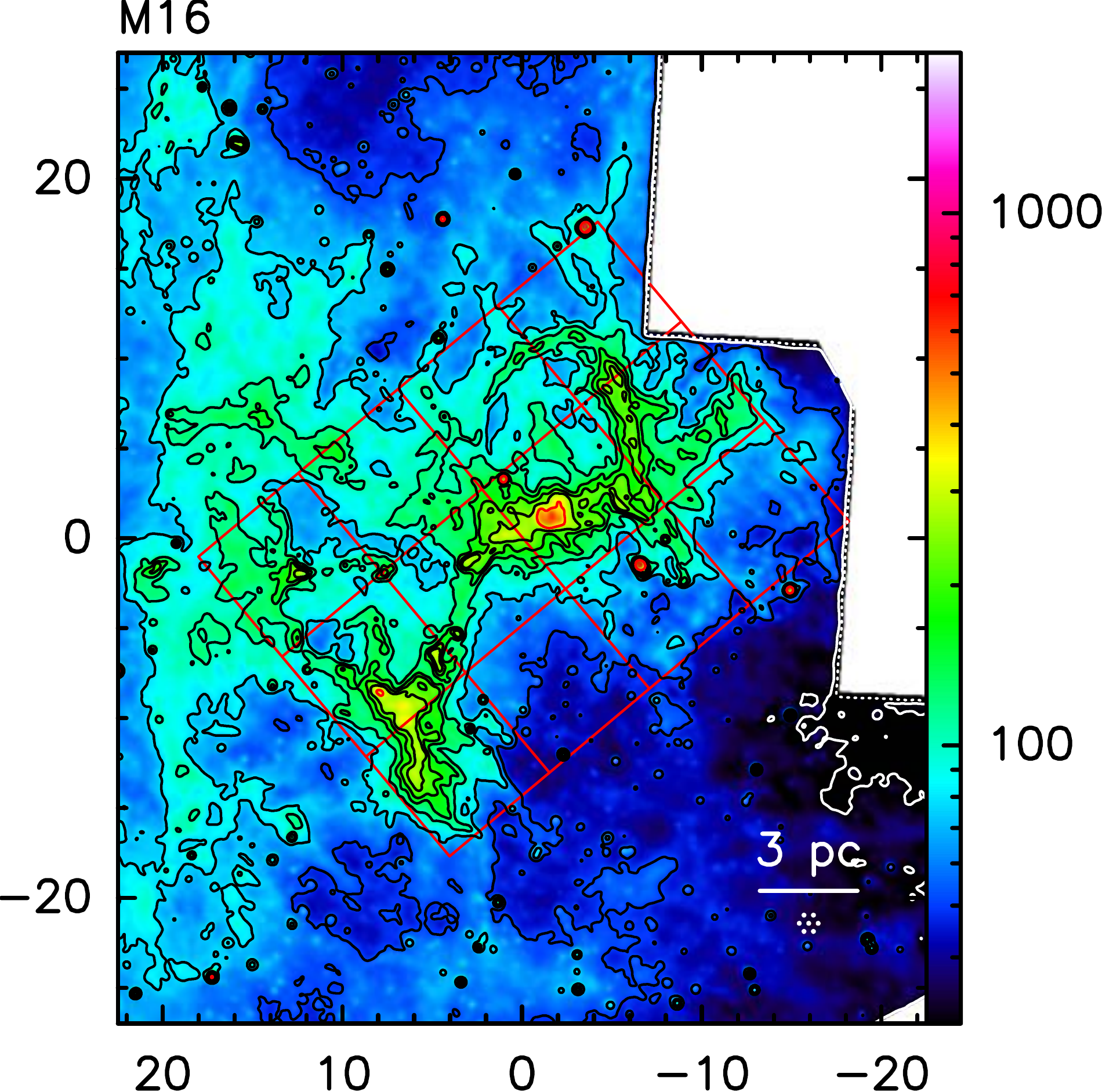} 
\includegraphics[angle=0,width=5.1cm,height=7cm,keepaspectratio]{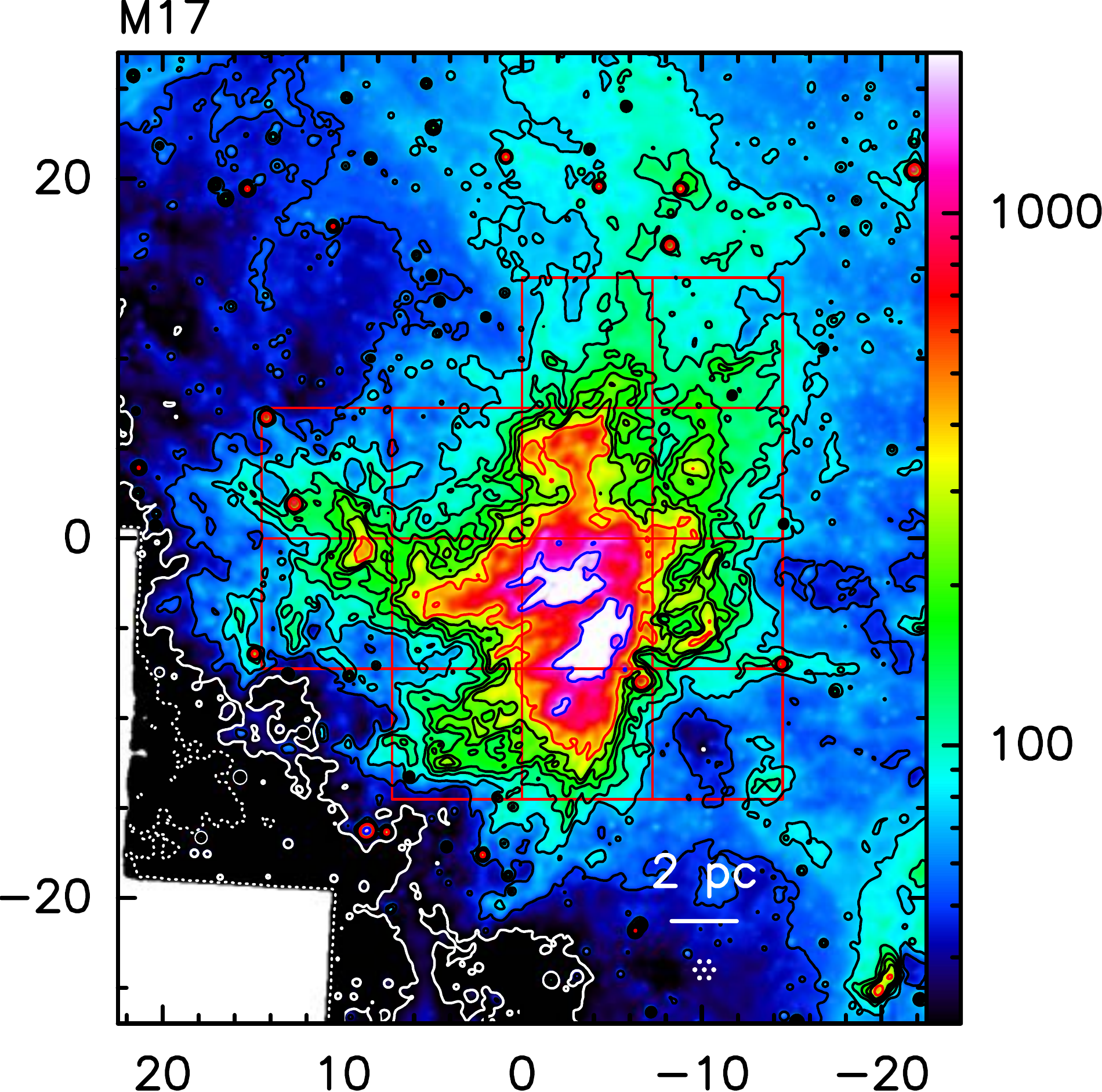} 
\includegraphics[angle=0,width=5.1cm,height=7cm,keepaspectratio]{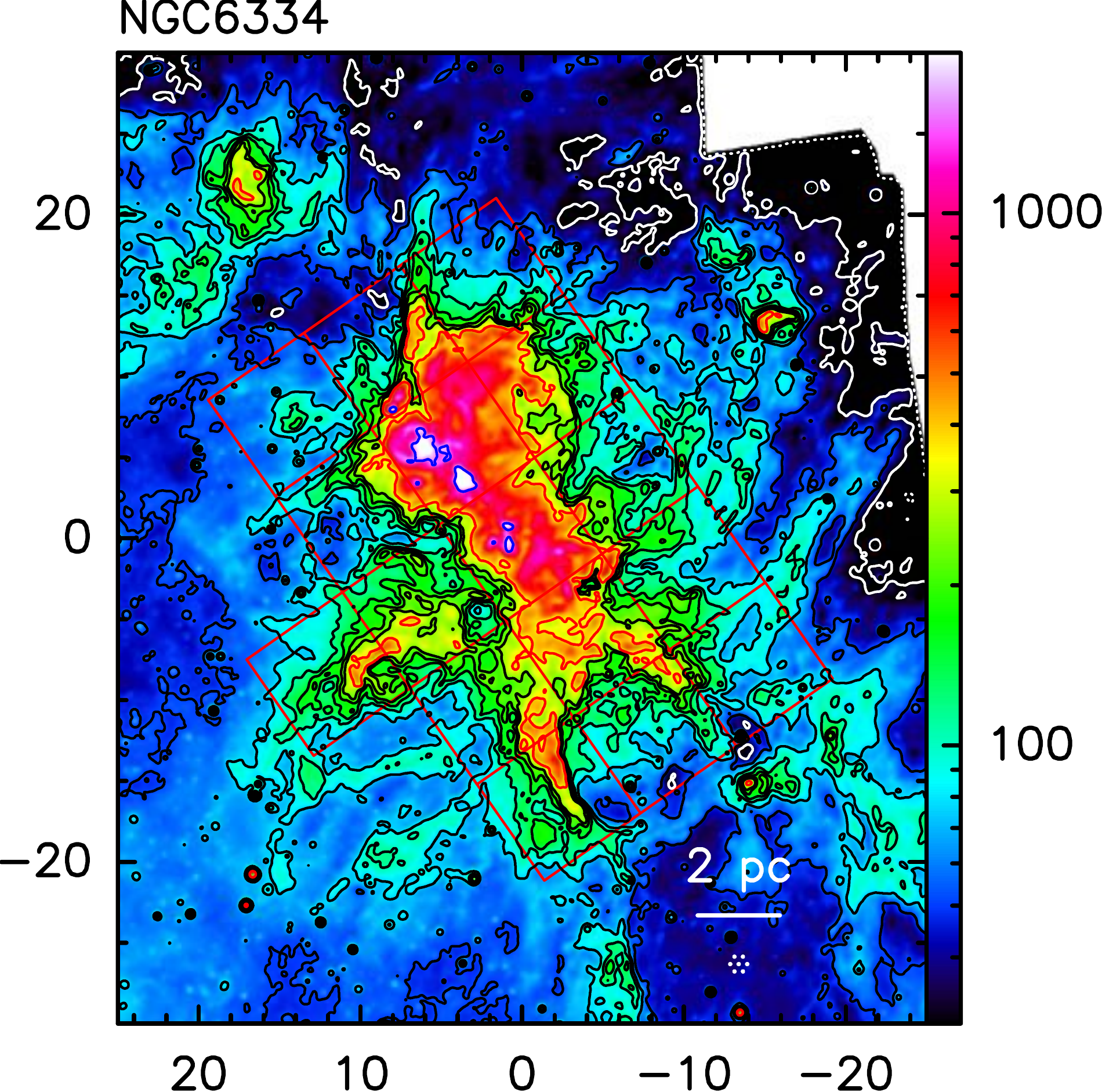} 
\includegraphics[angle=0,width=5.1cm,height=7cm,keepaspectratio]{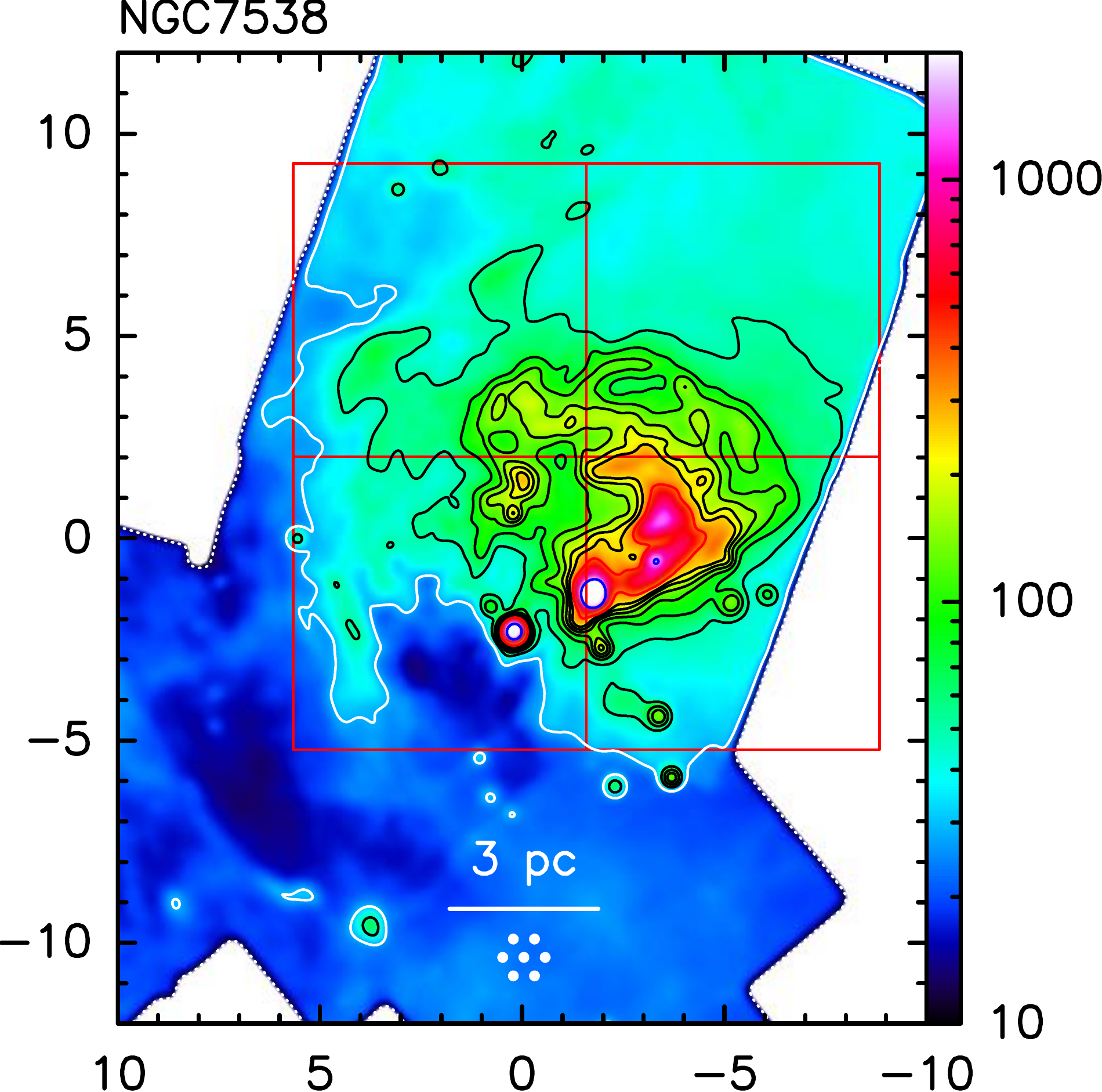} 
\includegraphics[angle=0,width=5.1cm,height=7cm,keepaspectratio]{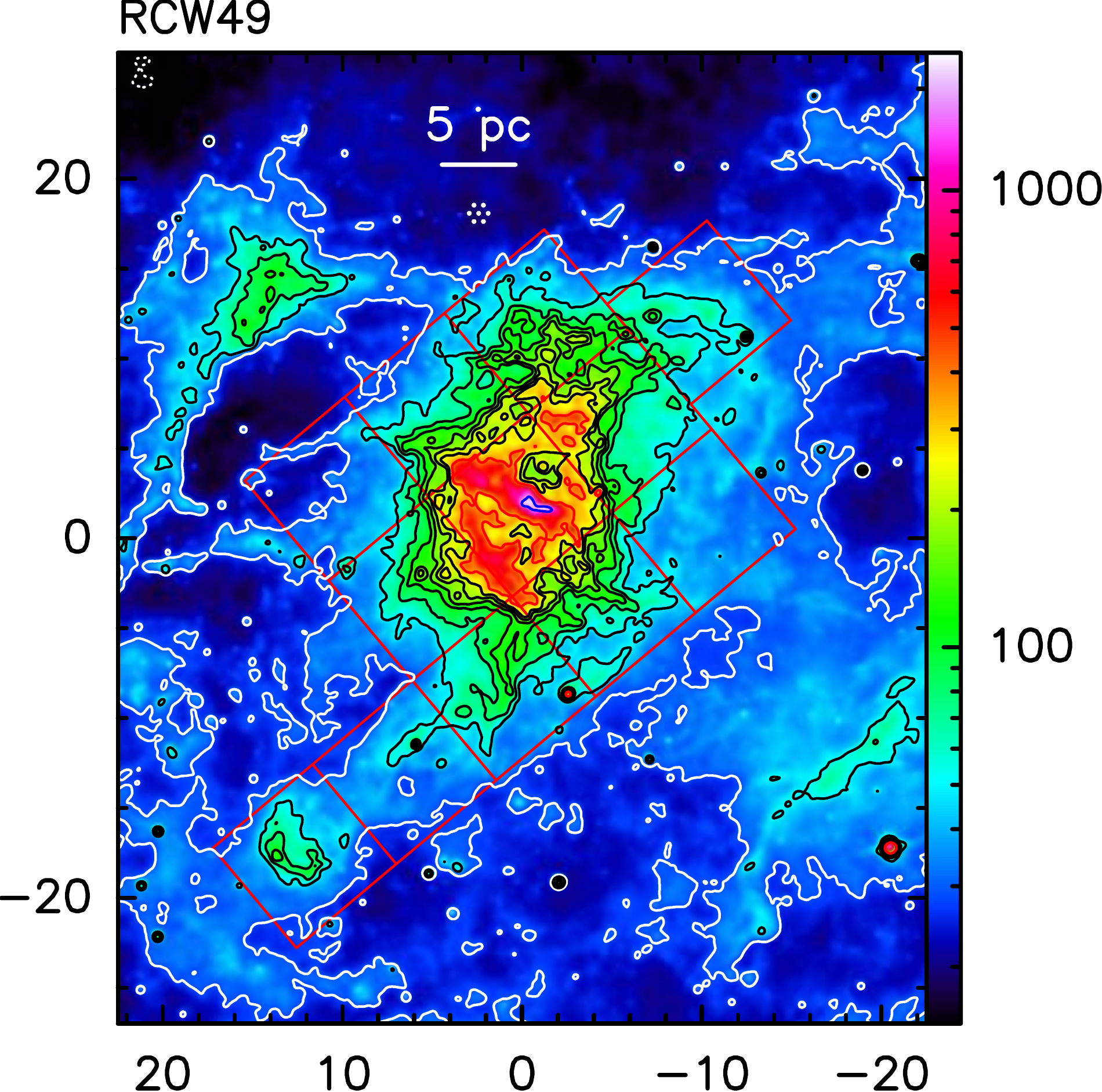} 
\includegraphics[angle=0,width=5.1cm,height=7cm,keepaspectratio]{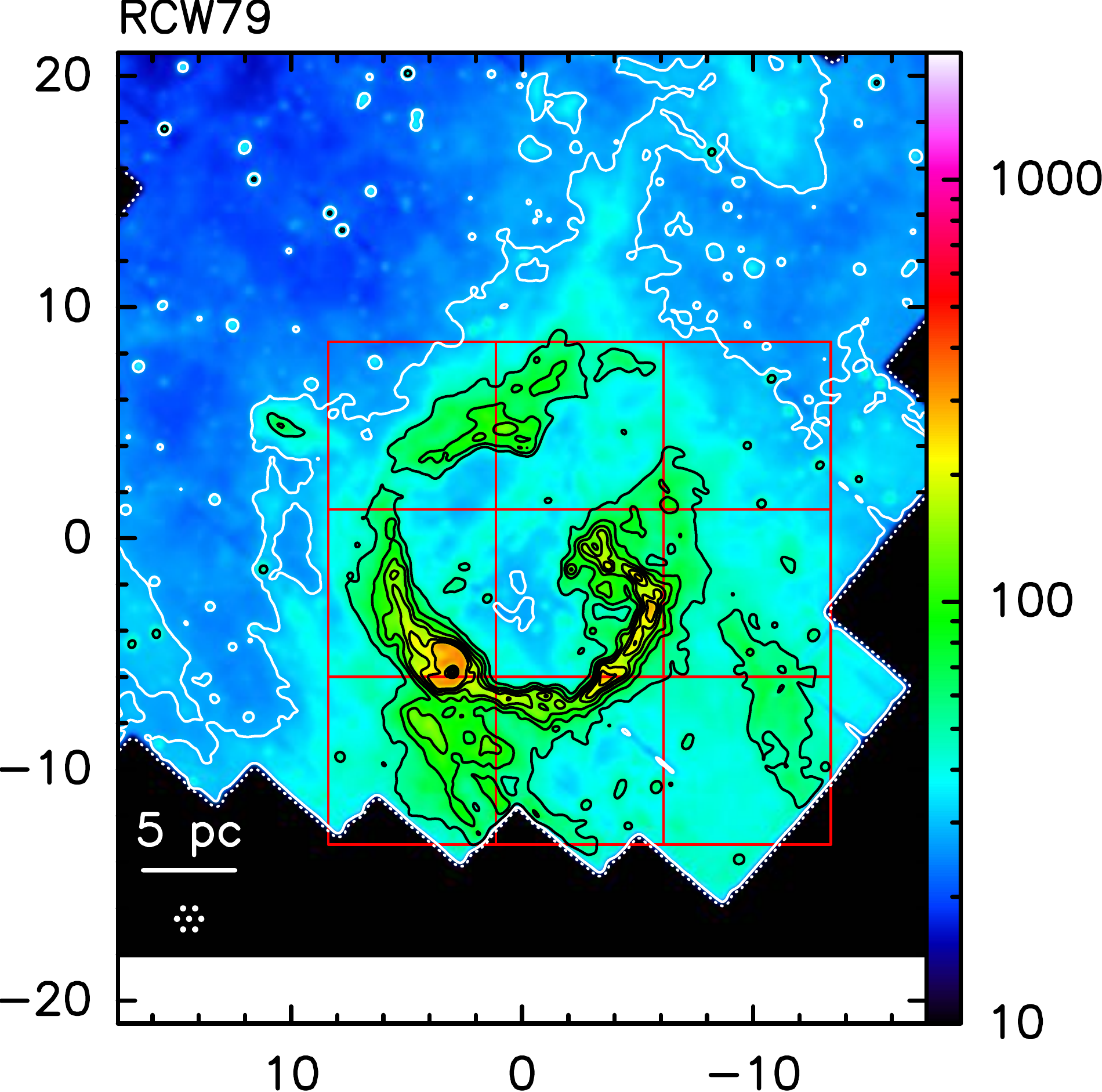} 
\includegraphics[angle=0,width=5.1cm,height=7cm,keepaspectratio]{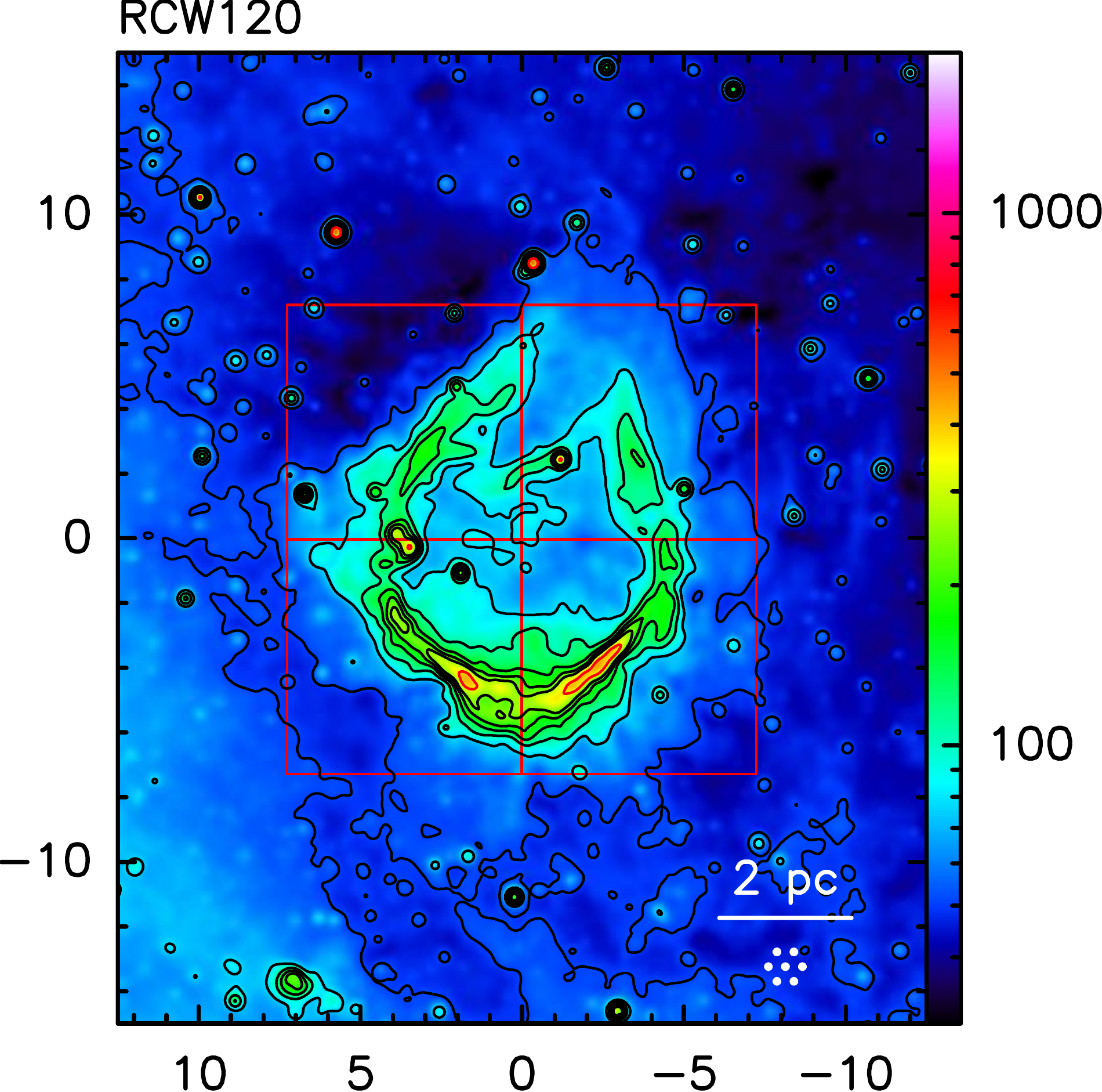} 
\includegraphics[angle=0,width=5.1cm,height=7cm,keepaspectratio]{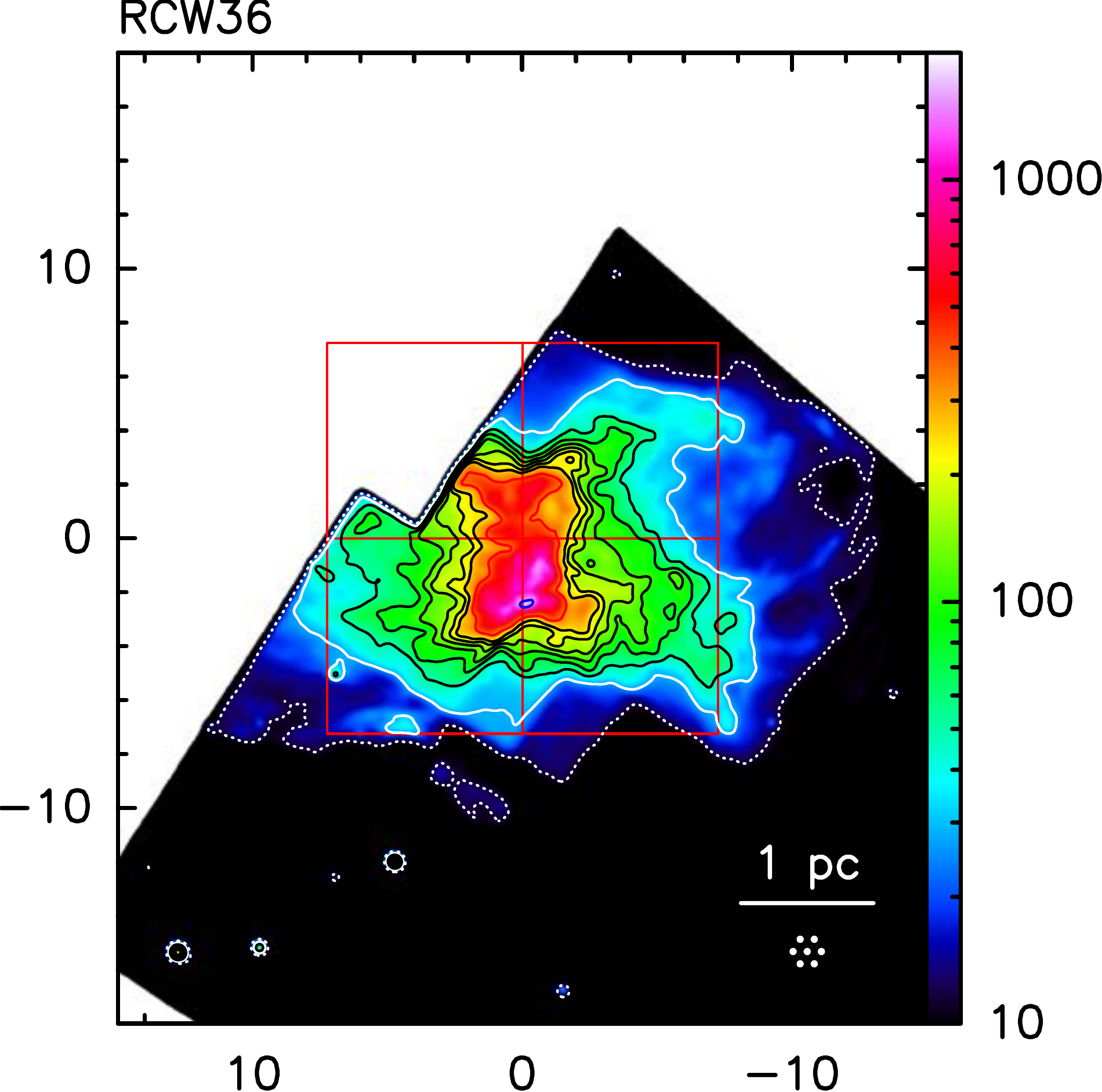} 
\includegraphics[angle=0,width=5.1cm,height=7cm,keepaspectratio]{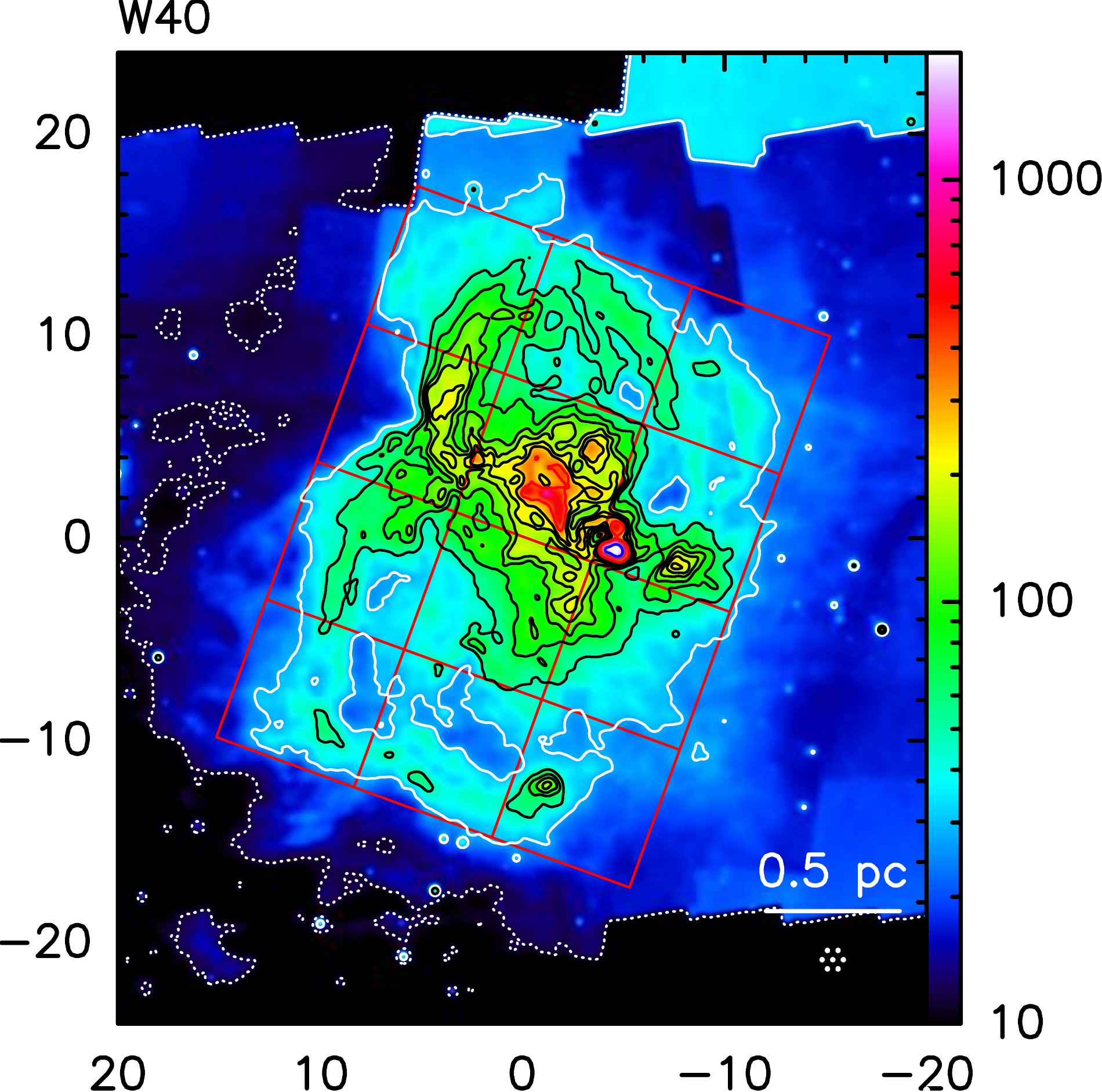} 
\includegraphics[angle=0,width=5.1cm,height=7cm,keepaspectratio]{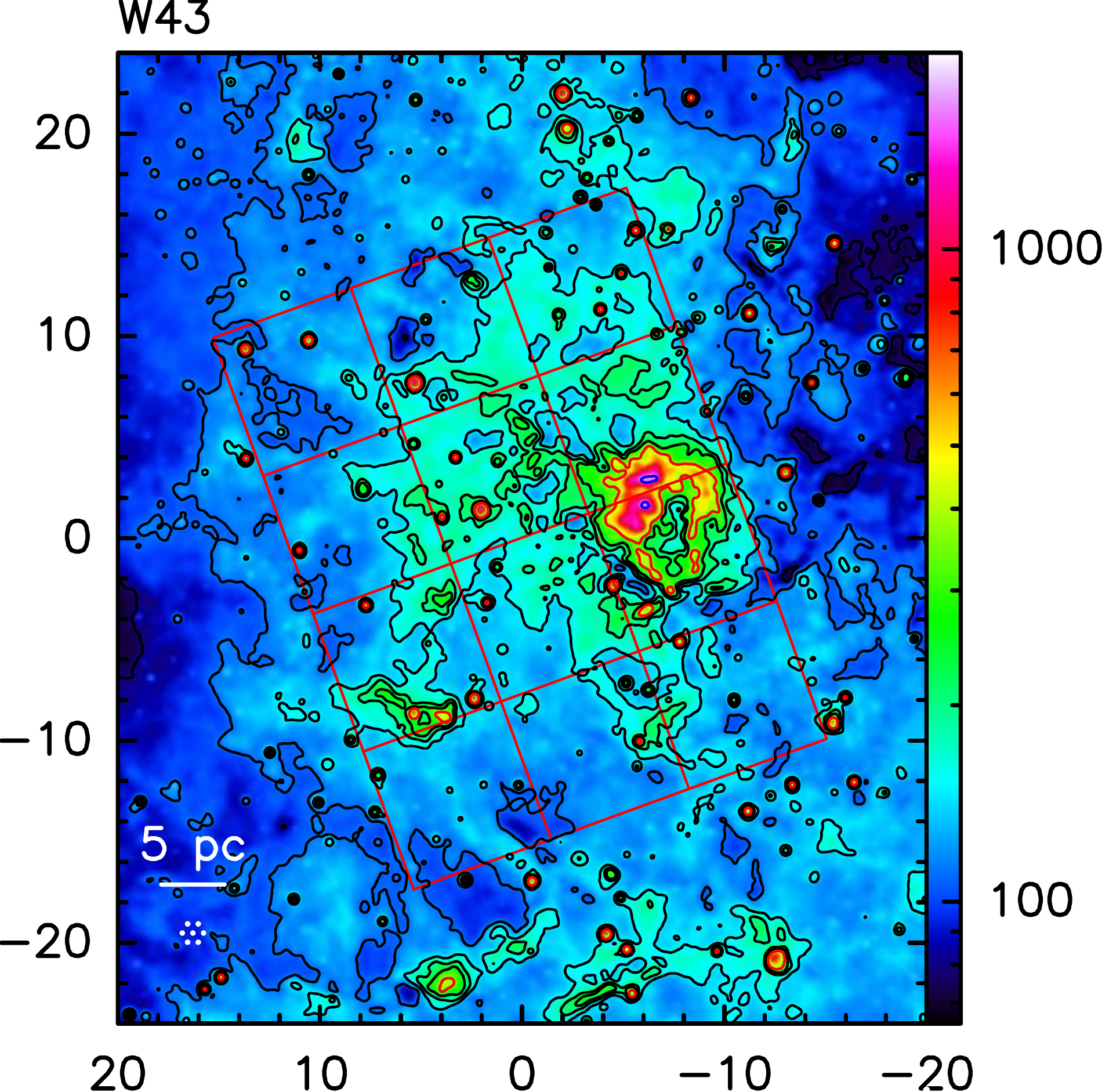} 
\end{center}
\caption[]{IRAC 8 $\mu$m maps of the FEEDBACK sources in color (in
  MJy/sr), convolved to the upGREAT beam of 14.1$''$. Contours are
  predicted \CII\ integrated line intensity based upon the \CII-8
  $\mu$m relation derived for L1630, Orion, and 30 Dor
  \citep{pabst2017,pabst2019}. Contour levels are: white dashed (50 K
  km s$^{-1}$), white (100 K km s$^{-1}$), black (150, (50), 400 K km
  s$^{-1}$), red (500 K km s$^{-1}$), blue (1000 K km s$^{-1}$).  The
  LFA/upGREAT 7 beam pattern is plotted in a corner in each box.  The
  overlaid red boxes represent the tiles that cover the \CII\ mapping
  area.  The images show RA, Dec offsets in arcmin with respect to the (0,0)
  position given in Table~\ref{table:summary1}. }  \label{sourcesfigure}
\end{figure*}

\section{Goals - FEEDBACK Key science} \label{goals}

Investigating the effects of stellar feedback from massive stars is a
vast science topic but we reduce it here to these questions: 1) How
large is the kinetic energy input into the ISM and, connected to this,
how does the surrounding medium react?  2) How does the radiative
coupling of interstellar gas to the FUV photons work and how does the
gas respond? \\ We specifically address these subjects by mapping the
\CII\ 158 \micron\ and \OI\ 63 \micron\ lines in Galactic massive
star-forming regions and, thus, focus in the following on how these
observations help to assess stellar feedback effects.

\subsection{Morphology and dynamics} \label{dynamic} 

Infrared (IR)- and FIR-surveys of the Milky Way, carried out with {\sl
  Spitzer, WISE}, and {\sl Herschel}, and radio surveys, have revealed
many thousand \HII\ regions
\citep{paladini2003,churchwell2006,deharveng2010,anderson2011,anderson2014}.
\citet{anderson2011} have shown that approximately half of all
\HII\ regions have a ring-like morphology, commonly referred to as
'bubbles'.  For the other \HII\ regions, their geometry varies between
bipolar structures and more complex regions with fragmented shells,
pillars, and globules\footnote{See also the 'The Milky Way Project'
  \citep{simpson2012} catalog of bubbles
  \citep{beaumont2014,jayasinghe2019}, based on {\sl Spitzer}.}.
Within FEEDBACK, we observe template regions for these different
geometries over a large range of size scales (from a few pc to tens of
pc).

In the classical view of \citet{spitzer1968}, expanding \HII\ regions,
excited by massive stars, sculpt ionized spherical cavities in their
surrounding molecular clouds. Because stellar radiation both ionizes
and heats gas up to $\sim$10$^4$ K, it creates a strong overpressure
that expands the gas into the surrounding cold molecular cloud. For
homogeneous initial conditions, this expansion can be described by a
simple analytical solution \citep{spitzer1968}. For very massive
stars, stellar winds are important as well since they drive shocks
into the surrounding medium, creating XDRs with temperatures up to
10$^4$~K in regions with high gas column density and - in the fully
ionized phase - a hot, but very tenuous, plasma with a typical density
of n$\sim$1~cm$^{-3}$ and temperatures up to 10$^6$~K.  The resulting
overpressure of this plasma leads to the formation of dense swept-up
shells on the surfaces of surrounding clouds \citep{weaver1977}.  The
inside of this shell is ionized by the stellar EUV photons, while the
shell forms a dense PDR. Energy conduction from the hot plasma to the
dense shell across the contact discontinuity that separates the two,
results in additional cooling, as well as mass loading of the
plasma. While this process is not well understood, it likely controls
the dynamics of the shell. Finally, radiation pressure on the dust and
gas assists the expansion, or even dominates under extreme conditions
\citep{draine2011,haid2016,krumholz2009}.  Note that stellar wind is
not the only way to have a shock because the expansion velocity of the
ionized gas into the surrounding medium is supersonic.  The shock
front precedes the ionization front as the expansion continues and the
surrounding material is locked into a layer (between the shock front
and the ionization front). In simulations \citep{dale2014}, the
wind-blown \HII\ regions resemble better observed \HII\ region
morphologies than those created by ionization alone.  More specific,
the relative effects of stellar winds are strongest at early times
when the massive star(s) are embedded in dense gas from the molecular
cloud.  Once this gas has been cleared away, the expanding
\HII\ region dominates the dynamical and morphological evolution of
the cloud. It is one of the main objectives of FEEDBACK to discern if
\HII\ regions are wind-driven or ionization-driven, or both, depending
on physical parameters such as density, radiation field and spectral
type(s) and ages(s) of the exciting star(s).

Each of these mechanisms leads to the creation of a dense shell of gas
and the density profile along a cut orthogonal to the ionization front
will be asymmetric, with a sharp gradient toward the
\HII\ region. Such profiles were indeed observed, using for example
{\sl Herschel} dust column density maps \citep{peretto2012,schneider2013,tremblin2013,tremblin2014}.
The PDR is typically trapped in this shell and its contribution will dominate the
\CII\ line emission. Hence, we can assess the distribution and the dynamics of
the warm PDR gas via the \CII\ line, while low- to mid-J CO lines
trace the molecular gas further away from the ionization front as well as the
surrounding unperturbed cold molecular cloud material, see, e.g.,
\citet{peng2012,mookerjea2012,stock2015,nagy2017,joblin2018,schneider2018}.
The high angular and spectral resolution data of the \CII\ line will reveal the
distribution and velocity field of the PDR material. Some of the 
\CII\ emission may originate in the the \HII\ region itself and this will
have to be distinguished through various diagnostics. First, the \CII\ emission
distribution, velocity field and line width need to be compared to other data sets, i.e., with
diffuse X-ray emission, molecular line data, dust column density and temperature maps. 
Position-velocity (pv) diagrams and channel-maps of \CII\ will show
signatures of near and far molecular walls if the bubbles are 3D
shells.  Evaluating the energetics, \CII\ observations will also show 
whether such a \CII\ bubble expands due to the impact of stellar winds or
the pressure-driven expansion of the \HII\ region. In addition, the contribution of
strong stellar winds can be assessed by X-ray emission. Several of our targets
have been observed by Chandra in the MOXC surveys and show diffuse X-ray emission, tracing stellar
wind activity \citep{townsley2014,townsley2018,townsley2019}.

Accompanying observations of CO and/or other molecular lines tracing
the molecular cloud will then show how such an expansion continues in
the cold and denser gas phase. However, disentangling the primordial
cloud structure and the effects of winds and photoionization carving
into the clouds will be a challenge (see also Sec.~\ref{triggering}).
Line profiles of \CII\ are fitted using multicomponent Gaussians to
determine surface brightness and identify coherent velocity and cloud
structures/components from the channel and pv maps. Because of the
high sensitivity of our data, we are able to detect the \tCII\ hfs
emission in many sources. This allows by comparison of the line
profiles study of whether the \CII\ line is affected by
self-absorption and thus might mimic unreal kinematics.

A recent \CII\ study of the Orion molecular cloud
\citep{pabst2019,pabst2020} displays clear observational signatures of
several expanding structures. The Veil nebula dominates the large
scale appearance of the region.  Over the last $\sim$100 000 years,
the stellar wind from the massive O7 star $\theta^1$ Orionis C has
blown a 4 pc diameter half sphere containing some 2000 M$_\odot$ of
gas expanding at 15 km s$^{-1}$ toward the low pressure region in
front of the OMC1 core. The cavity of the shell is filled by hot
plasma from the shocked stellar wind \citep{guedel2008}.  The total
kinetic energy of this expanding half-shell is comparable to the total
mechanical energy delivered by the wind. The \CII\ observations of the
Orion Molecular cloud also reveal expanding shells surrounding M43 and
NGC1977.  In these two cases, the overpressure in the ionized gas,
created by these B0.5 stars, gives rise to a Spitzer-type expansion
\citep{pabst2020}.

\subsection{Triggering of star-formation} \label{triggering}

The \CII\ observations form the basis for a study on the dynamical
response of newly formed stellar clusters and the disruption of their
nascent clouds.  Dense gas condensations, i.e., the locations of
future stars, might form as a result of fragmentation of the neutral
material due to the gravitational instability during the expansion of
the \HII\ region. These cold, dense cores in the
\HII\ region/molecular cloud interface of the FEEDBACK sources can be
traced by submm dust continuum observations, either already performed
by {\sl Herschel} \citep{tige2017,liu2017,zhang2020}, or using
ground-based instruments such as ArT\'eMiS
\citep{hill2012a,andre2016,zavagno2020} or LABOCA \citep{massi2019}
and - for detailed studies - ALMA and NOEMA.

So far, an overdensity of young stellar objects is observed at the
edges of \HII\ regions, and up to 25\% of the ionized regions show
high-mass star formation triggered on their edges
\citep{deharveng2010,thompson2012}.  In the picture of sequential
triggered star-formation \citep{elmegreen1977}, there should be an age
gradient observed in the spatial distribution of young sources in the
surrounding molecular cloud \citep{martins2010}. Furthermore,
radiative feedback will impact gas and dust temperatures within the
molecular cloud, modifying the initial conditions for collapse and
affecting proto- and young stellar objects in different evolutionary
phases.  However, observational studies of this issue are
inconclusive.  Taking the example of the Rosette Molecular cloud,
\citet{balog2007} reveal an increase of the average NIR excess
fraction for stellar clusters with distance to the cluster center, and
\citet{williams1994} show that star-formation activity is more intense
in the \HII\ region/molecular cloud interface region than in the
molecular cloud center. On the other hand, \citet{roman-zuniga2008}
and \citet{cambresy2013} found that the relative age differences of
the clusters are not consistent with a sequential triggered star
formation scenario. \citet{schneider2012a} argue that star formation
takes place in filaments and filament mergers that arose from the
primordial turbulent structure but can be locally induced in the
direct interaction zone between an expanding \HII\ region and the
molecular cloud.

The \CII\ data, combined with molecular observations, will provide
physical conditions, and in particular pressures in these regions and
will help to evaluate if triggered star formation can occur in dense,
swept up shells (Sec.~\ref{energy}).  This will be a difficult task
but we will for example investigate the variation of the \CII\ and
molecular line velocities and line widths from the interface region
into the surrounding cloud.  Clumpiness of the cloud needs to be taken
into account because it enables UV radiation to penetrate deeper into
the cloud, giving rise to \CII\ emission from many PDRs along the
line-of-sight. {\sl Herschel} data provide the distribution of the
dust temperature that will be compared to the gas temperature, derived
from CO.

\subsection{Kinetic and radiative energy} \label{energy}

The amount of energy and momentum that drive the dynamics of
\HII\ regions need to be evaluated for each source in order to assess
which mechanism is the dominant process for each source.
\citet{pellegrini2007,pellegrini2011,lopez2011,lopez2014} for example
provide a comprehensive overview how the pressure (or energy) terms
for direct radiation, dust-processed IR radiation, warm ionized gas,
and hot, shock-heated by stellar winds are determined
observationally. We here focus on how \CII\ and \OI\ observations,
together with complementary data, can be used for estimating the
energy budget.

The kinetic energy input into the ISM by massive stars will be
investigated as a function of star formation activity (cluster size,
spectral type, stellar wind) and evolutionary stage of the region both
in terms of the generation of large scale motions as well as the local
injection of turbulence into molecular clouds. The \CII\ observations
reveal the morphology of expanding structures, and their masses are
determined from the \CII\ column density and compared to the molecular
gas mass, estimated from {\sl Herschel} dust column density maps and
CO maps.  The \CII\ line can be affected by high optical depth, so
that mass estimates are lower limits, but observations of the
optically thin \tCII\ hfs emission can be employed. The integrated
intensity of the \tCII\ hfs emission is expected to be only a few
percent of the \twCII\ emission (typically 50 times lower),
depending on the elemental $^{12}$C/$^{13}$C abundance ratio and the
optical depth of the emission in the main isotope. However, averaging
over a large area, the Signal-to-Noise ratio (S/N) of the \tCII\ hfs emission can be
sufficiently increased to derive the optical depth and hence a lower
limit on the mass of emitting gas. For shells expanding towards us,
extinction measurements of the stars in the ionized gas can also
provide an estimate of the column density through the shell and,
hence, its mass. Combined with the measured expansion velocity, this
yields the kinetic energy of the gas, which can be directly compared
to the thermal energy of any hot plasma present (determined from X-ray
observations) and of the ionized gas. The ionizing photon flux and the
total luminosity of the region are estimated from optical, radio, and
IR observations. The spectral types of the ionizing stars in each
cluster have been determined directly through infrared
spectroscopy. Hence, we can observationally link the kinetics of the
shell to the stellar characteristics (wind mechanical energy,
luminosity).

We can also quantify the radiative coupling of PDR gas to the FUV
photons. \CII\ line intensities are directly compared via correlation
plots to CO emission, IRAC 8~$\mu$m and WISE 12~$\mu$m emission due to
PAHs, and PACS 70, 160~$\mu$m (and 100 $\mu$m if available) far-IR
dust emission, all convolved to the same beam size. The \CII\ line is
the dominant cooling line of the gas and by comparing the integrated
line intensity to the total infrared dust and PAH emission we
empirically derive the heating efficiency in terms of the \CII/FIR (or
\CII/TIR\footnote{TIR is the total infrared flux between 3 and
  1100~$\mu$m \citep{dale2002}, while FIR is the total FIR flux
  between 42.5 and 122.5~$\mu$m \citep{helou1988}.}), \CII/CO and
\CII/PAH emission ratios (c.f.,
\citet{okada2013,pabst2017,anderson2019}).  Theoretically, the heating
efficiency depends on the ionization parameter,
$\gamma$=G$_0T^{1/2}/n_e$ with G$_0$ the intensity of the radiation
field in terms of the average interstellar radiation field, $T$ the
gas temperature and $n_e$ the electron density \citep{bakes1994}. As
PAHs and grains charge up (large $\gamma$), the ionization potential
increases and fewer photons can further ionize the
PAHs/grains. Moreover, the photoelectron has to overcome an increased
Coulomb potential, diminishing the energy delivered to the gas. Hence,
in a stronger radiation field or lower density region, PAHs and very
small grains charge up and the heating efficiency drops. Presently,
there is limited data available \citep{okada2013,pabst2017}, showing a
decrease with ionization parameter as theory predicts. However,
overall, theory seems to overpredict the heating
efficiency. Preliminary analysis of the Orion data reveals a large
spread in heating efficiency at any given ionization parameter. These
variations seem to be linked to the spatial location in the region,
indicating a dependence on local conditions not caught by the
ionization parameter, or on the past evolution of the region.

FEEDBACK also has the potential for follow-up studies of $\gamma$-ray
emission from highly dynamic and ionized regions of the ISM. Several
of our targets show energetic $\gamma$-ray emission, detected with
\textit{Fermi} and/or HESS. These are Cygnus X, observed with
\textit{Fermi} \citep{ackermann2011}, RCW49 with HESS
\citep{aharonian2007,abramowski2011}, and W43 with \textit{Fermi}
\citep{lemoine2011} and HESS \citep{chaves2008}. M16, M17, and RCW36
are listed in the \textit{Fermi} Large Area Telescope Fourth Source
Catalog (4FGL, \citet{abdollahi2020}.

\subsection{Photodissociation regions} \label{pdr}

The coupling of FUV photons to the gas is a key process that sets the
structure and physical conditions in PDRs and their emission
characteristics \citep{hollenbach1999}, and regulates the phase
structure of the ISM \citep{wolfire1995,wolfire2003}.  The FEEDBACK
program allows in depth studies of the structure and characteristics
of PDRs on small ($\sim$1 pc) and large (tens of pc) physical scales.
The \CII\ line is also an important coolant of warm neutral atomic and
CO-dark \citep{wolfire2010} molecular gas in PDRs and the cloud
selection spans a wide range of physical conditions that are well
probed by \CII.  The (undersampled) \OI\ 63 $\mu$m maps will be bright
in the densest regions illuminated by strong FUV fields. We are also
obtaining complementary data in other PDR tracers, namely the
\CI\ line at 490 GHz, using the 4GREAT receiver on SOFIA and the APEX
telescope, and mid- to high-J CO lines using APEX.  We sample a wide
range of physical conditions including the radiation field (G$_0$
ranging up to a few times 10$^5$), (electron) density (typically from
$10^2$ to $10^6$ particles cm$^{-3}$), gas temperature (from $10^2$ to
$10^3$ K), and spectral type of the illuminating star (O9 to O4 and
WR).

In order to enable a more efficient way to compare \CII\ and \OI\ maps
with PDR models, we are upgrading the PDR Toolbox
\citep{pound2008,kaufman2006} at
\href{http://dustem.astro.umd.edu/pdrt}{http://dustem.astro.umd.edu/pdrt}.
The PDR code used to generate its underlying database of line
intensities has improved physics and chemistry. Critical updates
include those discussed in \cite{neufeld2016}, plus photorates from
\cite{heays2017}, oxygen chemistry rates from \cite{kovalenko2018},
and \cite{tran2018}, and carbon chemistry rates from
\cite{dagdigian2019}. We have also implemented new collisional
excitation rates for \OI\ from \cite{lique2018} (and Lique private
communication) and have included $^{13}$C chemistry along with the
emitted line intensities for \tCII\ and $^{13}$CO.  The new PDR
Toolbox covers many more spectral lines and metallicities and allows
map-based analysis so users can quickly compute spatial images of
density $n$ and radiation field G$_0$ from map data.  It has been
rewritten in Python and provides Jupyter notebooks for data analysis.
It also can support other PDR model codes such as KOSMA-$\tau$
\citep{roellig2006}, enabling comparison of derived properties between
codes.

KOSMA-$\tau$ differs from all other numerical PDR codes in that it is
simulating the PDR emission from a spherical model cloud. This allows
to model a wider range of astrophysical scenarios compared to
plane-parallel geometries. A superposition of many small clumps can
represent the inhomogeneous structure of the interstellar medium,
while the limit of large clumps approaches the plane-parallel picture.
\citet{stutzki1998} showed that the fractal nature of the interstellar
medium can be represented by an ensemble of clumps following a
well-defined size distribution. This allows us to successfully model
the highly structured PDRs in our observational sample by the
superposition of individual spherical model clouds. KOSMA-$\tau$ was
compared against other models in a dedicated benchmark study in 2007
\citep{roellig2007} and the code is continuously updated and improved as
new data (e.g., reaction rates, atomic and molecular data) become
available. Recent updates particularly refined the chemistry module of the code, including: \\

\noindent$\bullet$ Flexible inclusion of isotopomeric chemistry.  \\
\noindent$\bullet$ Improved treatment of linear and cyclic isomers of a given molecule. \\
\noindent$\bullet$ Inclusion of exothermal reaction energies as heating terms in the local energy balance.\\
\noindent$\bullet$ New solution algorithms in order to improve numerical stability.\\
\noindent$\bullet$ Introduction of full surface chemistry.\\
\noindent$\bullet$ Inclusion of time-dependent numerical solvers.\\

A detailed description of the current state of the code is in
preparation (R\"ollig et al. 2020, in preparation). Presently we are working
to incorporate non-local transport terms such as diffusion and advection
into the code framework. Moreover, we expand the three-dimensional
modeling capabilities of KOSMA-$\tau$-3D
\citep{cubick2008,labsch2017} and include the full continuum radiative
transfer in our simulations.

Within the FEEDBACK project, clumpy KOSMA-$\tau$ results are compared to
the plane-parallel results from the PDR Toolbox to test the diagnostic
power of the models given the spatially and spectrally resolved data
available. 

Energy can be simultaneously injected in radiative (FUV photons) and
kinetic (shocks) form.  In the regions with evidence for shocked gas
emission (for example large observed linewidths of \CII\ and/or \OI),
we will compare our data to the results of dedicated modeling of
irradiated shocks \citep{godard2019,lee2019}.  These models include
the effects of FUV photons on the physical conditions and chemistry of
the gas. For moderate-velocity molecular shocks (up to a shock
velocity of about 30~km~s$^{-1}$), the most recent reference model was
presented in \citet{godard2019}. Given the high number of input
parameters, the comparison with observations is only meaningful if the
number of observables (emission lines) is sufficient. In FEEDBACK
regions, where this is the case and the observations in terms of
linewidths justify it, our models will enable us to quantify the
effects of kinetic vs. radiative energy (see e.g. Figure 12 of
\citet{godard2019}). A large grid of models has been run in
preparation for the interpretation of FEEDBACK data, thoroughly
exploring input parameters such as the pre-shock density, the shock
velocity, the external radiation field, the magnetic field strength,
and the PAH abundance in the observed regions. A study is in
preparation to extend this model toward higher-velocity shocks (up to
60~km~s$^{-1}$). In such shocks, additional FUV photons are generated
by the shock itself (see the earlier works of \citet{hollenbach1989}).

Summarizing, our project will investigate issues such as the
relative importance of the \CII\ and \OI\ for gas cooling in regions
of different physical conditions, the role of self-absorption in
\CII\ and \OI\ emission and its effect on the analysis of low
resolution, Galactic and extragalactic observations, and the origin of
the \CII\ deficit in (Ultra)Luminous IR Galaxies.

\subsection{Formation of filaments, pillars, and globules} \label{ism}

Massive stars have a profound influence on the overall structure of
the ISM and the proposed observations can address such key questions
as: Are dense filaments formed due to the interaction of expanding,
compressed shells around \HII\ regions (see Sec.\ref{triggering})? How
is stellar feedback impacting an inhomogeneous molecular cloud or
cloud surface? Early studies using the \CII\ line
\citep{stutzki1988,meixner1992,schneider1998} have revealed that FUV
radiation can penetrate deep ($>$10 pc) into clouds with high clump to interclump
density contrast, thus inducing emission from multiple PDRs along the
line of sight.  Simulations have also shown that geometry plays an 
important role.  \citet{walch2012, walch2013,walch2015} demonstrated 
that for low fractal dimensions of the cloud, the border of the
\HII\ region is dominated by shells that break up into massive high-density clumps
('shell-dominated' region) while high fractal dimensions lead to the formation of many pillars
and cometary globules, containing compact dense clumps ('pillar-dominated' region). 
\citet{tremblin2012a,tremblin2012b} showed that UV radiation creates a
dense shell compressed between an ionization front and a shock ahead
and that density modulations in the interface produce a curved shock
that collapses on itself, leading to pillar-like structures that can
evolve into globules. 

With the \CII\ and \OI\ mapping, we are able to study on a small scale
($<$1 pc up to a few pc) all structures in our sample that are  
produced under the influence of radiation such as pillars, globules, proplyds,
and evaporating gaseous globules (EGGs).
This includes the famous 'Pillars of Creation' in M16 \citep{hester1996} as
well as less known, yet equally interesting, features in other regions. The velocity resolved
extended \CII\ and \OI\ maps are of use to investigate in more detail
the dynamics of pillars and globules and allow the detection of
velocity gradients, high-velocity outflowing gas and rotation. The
spatial emission distributions trace the external PDR surfaces as
well as internal heating sources, similar to what was found by
\citet{schneider2012b} for a globule in the Cygnus X region. The observed line
intensities and ratios will be applied to the KOSMA-$\tau$ model
\citep{roellig2006} and the PDR Toolbox for modeling the
FUV-irradiated regions of the pillars and globules in order to derive
the physical conditions. These can then be compared to dynamical
models of pillar and globule formation
\citep{williams2001,mizuta2005,gritschneder2009,miao2009,gritschneder2010,tremblin2012a,tremblin2012b}.

\section{Source selection} \label{sources}

\subsection{General considerations} \label{sources-general} 
We have selected a sample of sources that were included in the
{\sl Spitzer}/GLIMPSE and {\sl Herschel} HOBYS \citep{motte2010}, Gould Belt
\citep{andre2010} and Hi-Gal \citep{molinari2010} Legacy/Key
Programs. Northern sources are observed with flights originating from the
Armstrong Flight Research Center in Palmdale, California, and southern sources
with flights from Christchurch, New Zealand.  Table~\ref{table:summary1} and
\ref{table:summary2} and Figure \ref{sourcesfigure} give an overview of the
sample.  In the selection we ensured that a wide range in star
formation activity is covered, from regions dominated by single O
stars, by small groups of O stars, by compact clusters, by super star
clusters, and by mini starbursts\footnote{We follow the definition by
  \citet{motte2003} that a Galactic mini starburst region is a cloud
  with a star-formation efficiency as high as 25\%.}. Another
consideration was to include regions that are probably at different
stages of their evolution (see individual source descriptions
below). Morphology was an important criterion (see Sec. \ref{dynamic})
and we incorporated \HII\ regions that are (almost) perfect spherical
bubbles, multi-bubbles or broken bubbles, bipolar structures, all with
or without pillars and globules.  Regions with more dispersed star
formation activity and clouds with many filaments and massive ridges
were also considered. Furthermore, the selected sources include
regions dominated by the thermal expansion of ionized gas
(Spitzer-type expansion), by stellar wind driven flows, by radiation
pressure, by the concerted interaction of multiple expanding
\HII\ regions, by the presence of nearby rich OB associations, and by
the action of converging flows associated with the large scale spiral
arm structure of the Milky Way. This allows us to study feedback on a
wide range of scales both in size (from sub-pc to tens of pc) and
energy (energy of a single O star to that of clusters of stars
containing tens to hundreds of OB stars).

The sample comprises well-known sources dominated by these
different dynamical processes (thermal, wind, radiation pressure) and
covers the parameter space of star formation activities and
evolutionary stages. 

\subsection{Complementary data sets} \label{sources-complement} 

Much ancillary data is available for all regions as they have been
studied in depth in a variety of {\sl Herschel} and {\sl Spitzer}
programs. These surveys provide detailed spectral energy
distributions, luminosities, dust temperatures, and column densities
at spatial scales that are comparable to those that are obtained by
the upGREAT mapping of \CII\ and \OI. Source catalogs from various
instruments are available for a census of ongoing star formation
activity (pre- or protostellar cores, protostars).

We use existing molecular line data (mostly CO) from publically
available large surveys (e.g., the FUGIN project from the Nobeyama
telescope for M16, M17, and Aquila, and dedicated molecular line and
atomic carbon observations using the APEX telescope
\citep{guesten2006}.  We already obtained $^{12}$CO and $^{13}$CO
3$\to$2 maps for some sources, performed recently with the new LAsMA
array on APEX (RCW120 was amongst these sources and spectra are shown
in Sec.~\ref{rcw120}). LAsMA is a 7-pixel single polarization
heterodyne array that allows simultaneous observations of the two
isotopomers in the upper ($^{12}$CO) and lower ($^{13}$CO) sideband of
the receiver, respectively.  The array is arranged in a hexagonal
configuration around a central pixel with a spacing of about two beam
widths (the beamwidth is 18.2$''$ at 345.8 GHz) between the pixels. It
uses a K mirror as de-rotator. The backends are advanced Fast Fourier
Transform Spectrometers \citep{klein2012} with a bandwidth of
2$\times$4 GHz and a native spectral resolution of 61 kHz.  APEX will
also be used for observations of the atomic carbon line at 490 GHz and
the CO 6$\to$5 line.

The \CI\ line at 490 GHz was already mapped with the 4GREAT instrument
on SOFIA for NGC7538 and Cygnus (in parallel, 3 high-J CO lines were observed).
More observations of \CI\ and CO lines and of the \OI\ 145 $\mu$m line
for the FEEDBACK sources are planned. 

We also recently finished a program to observe radio-recombination lines
of hydrogen, helium, and carbon in M17, M16, W40, Cygnus X, and
NGC7538 at the Green Bank Telescope. 

\subsection{Details of individual sources} \label{sources-detail} 

Information on the individual sources (and of the program in general)
is found on the webpages of the FEEDBACK program:
\href{https://feedback.astro.umd.edu}{https://feedback.astro.umd.edu}
and
\href{https://astro.uni-koeln.de/18620.html}{https://astro.uni-koeln.de/18620.html}.
We here give a short summary for each source (see also Table~\ref{table:summary1} and
\ref{table:summary2}). \\

\noindent{\bf Cygnus X} \\
The Cygnus X region \citep{reipurth2008} is one of the richest star
formation sites in the Galaxy, mainly excited by the Cygnus OB2
association that contains well over 100 OB stars
\citep{comeron2002,wright2015}.  Most of the molecular clouds are
located at a distance of 1.4~kpc \citep{rygl2012}, the total molecular
gas mass is a few 10$^6$ M$_\odot$ with an average density of $\sim$60
cm$^{-3}$ \citep{schneider2006} and densities $>$10$^5$ cm$^{-3}$ in
clumps and cores associated with ongoing star-formation.  The northern
part of Cygnus X contains the prominent DR21 and W75N regions that are
located within dense filamentary structures, called 'ridges'
\citep{schneider2010,hennemann2012}. The average UV-field is high
($\sim$300 G$_\circ$) and reaches peak values up to $\sim$10$^5$
G$_\circ$ in PDRs close to Cyg OB2 \citep{schneider2016}. The
interaction of UV radiation with the molecular clouds created a wealth
of structures such as pillars, globules, EGGs, and proplyd-like
features (proplyds are evaporating circumstellar disks, but the
sources found by \citet{wright2012} in the immediate environment of
the Cyg OB2 association are much larger than typical proplyds). Based
on this classification using {\sl Herschel} data,
\citet{schneider2016} proposed an evolutionary scheme in which pillars
can evolve into globules, which in turn then evolve into EGGs,
condensations and proplyd-like objects.

We study the northern Cygnus X region because it shows many sequential
stages of star formation and - as a large complex - is a key to
understanding the role that massive star complexes play in external
galaxies. The Cygnus X North region is exposed to a high overall
radiation field, mostly arising from Cyg OB2, but also from many local
\HII\ regions. For FEEDBACK, we focus on mapping the DR21 ridge and
the 'Diamond Ring' \citep{marston2004}, an \HII\ region southwest of
DR21. \\

\noindent{\bf M16} \\
The Eagle Nebula (M16) is a young (1-3$\times$10$^6$ yr), active
high-mass SF region in Serpens. \citet{guarcello2010} proposed that
the stars in the northwest part of the \HII\ region are younger than
the stars in the southeast part and that a 200~pc shell triggered the
formation of both M16 and M17 3 Myr ago on much larger scales.
Responsible for heating and ionizing the Nebula is the young open star
cluster NGC6611, containing four early-type O stars, leading to a high
average UV-field of $\sim$300~G$_\circ$. The transition between
\HII\ region and dense, cold gas is rather sharp with many
UV-illuminated features. In particular Hubble Space Telescope imaging
of the 'Pillars of Creation' has made M16 iconic.

M16 is well suited to study the mechanical (stellar winds) and
radiative energy (UV field) incident on the molecular cloud complex,
shaping the gas into pillars and other features. In contrast to the
Cygnus X region, M16 does not show free-floating objects such as
globules or proplyds.  We thus focus on pillar structures that
are a common morphological phenomenon, appearing on the boundaries of
many evolved \HII\ regions \citep{dent2009,pound2003,xu2019}. 

{\sl Herschel} studies \citep{hill2012b} showed that the cluster
affects the temperature within the molecular cloud, modifying the
initial conditions for collapse and affecting the evolutionary
criteria of protostars, for example increasing the bolometric
temperature and the $L_{submm}/L_{bol}$ ratio.  The brightest area of
M16 is smaller than other massive SF regions and can be fully covered
in the \CII\ line, including the Pillars, the Spire, and the Arch. \\

\noindent{\bf M17} \\
M17 is located at a distance of 1.98~kpc \citep{wu2019} and associated
with the highly obscured (A$_V>$10) cluster NGC6618 with more than 100
OB stars.  The edge-on geometry allows to study feedback of the rich
cluster with the molecular cloud all along the interface and into the
highly clumpy molecular cloud. The total mass of the complex (from CO
observations) is 2$\times$10$^4$ M$_{\odot}$ and can be split into M17
North (M17-N) and M17 Southwest (M17-SW).  For M17-N (M17-SW), the
density ranges between 10$^{4}$-10$^{5}$ cm$^{-3}$ (10$^6$ cm$^{-3}$)
\citep{stutzki1990,meixner1992,wilson2003}. M17-SW has been studied
extensively in many different tracers and frequencies.  A recent SOFIA
[$^{13}$C\,{\scriptsize II}] study reveals unexpectedly large columns
of warm and cold \CII\ \citep{guevara2020}. \\

\noindent{\bf NGC6334} \\
NGC\,6334 is a very active star forming region (see e.g.,
\citet{persi2008}) with a remarkably large number of \HII\ regions
spread across the complex (7 compact and optical \HII\ regions per
square degree) at a distance of 1.3~kpc \citep{chibueze2014}. Most of
the \HII\ regions have a bubble-like morphology in the IR, FIR and
optical, but there are also examples of an expanding wind shell-like
\HII\ region (GUM61) or champagne flow (GUM64b), see e.g.,
\citet{russeil2016}.  The associated molecular cloud consists of a
$\sim$10~pc long, dense filament that is associated with strong
extinction and is embedded in a larger $\sim$50~pc-long, less dense
filamentary cloud \citep{zernickel2013,russeil2016}. A number of
active star-formation sites exist within the dense filament
\citep{brogan2016,tige2017,juarez2017,sada2020} with compact
\HII\ regions detectable at radio wavelengths such as NGC\,6334-I and
NGC\,6334-E being driven by a rich embedded cluster of B-type
stars. The average UV-field ($\sim$580~G$_0$) is above the median
within the selected sample.

The gas dynamics in NGC\,6334 are dominated by the large number of
\HII\ regions \citep{russeil2016}, in particular for the dense 10~pc
long filament. This one is also undergoing longitudinal collapse
\citep{zernickel2013} and is also likely compressed by the expansion
of the two large \HII\ region bubbles to the north and south
\citep{russeil2013}.  Structure analysis based on the
$\Delta$-variance method applied to the \textit{Herschel} H$_2$ column
density maps identified characteristic scales that can be caused by
the injection of energy due to expanding \HII\ regions
\citep{russeil2013}. \\

\noindent{\bf NGC7538} \\
NGC7538 is a small \HII\ region, which is part of the Cas OB2 complex
at a distance of 2.65 kpc \citep{moscadelli2009}. It is illuminated by
a group of O stars of spectral type O6-O9. The dominant ionizing star
of NGC7538 has been classified as an O3 or an O5 star
\citep{puga2010,ohja2004}, the radio data is consistent with an O5
star \citep{luisi2016}. \\
NGC7538 is expanding into a massive star forming cloud south and
southeast of the \HII\ region and has most likely triggered star
formation in the cloud. Especially the extremely young O star IRS1,
which is located at the boundary between the expanding \HII\ and the
molecular cloud, is still heavily accreting and surrounded by a
cluster of mm-continuum sources, which are probably all young
pre-main-sequence sources \citep{frau2014}. \\

\noindent{\bf RCW49} \\
RCW49 is one of the most luminous and massive \HII\ regions of the
southern Galaxy. From recent GAIA observations, its distance is
determined to be 4.21 kpc \citep{cantat2018}, slightly further away
than the 4.16 kpc determined by \citet{vargas2013} from
spectroscopy. It hosts the Westerlund 2 (Wd2) cluster, comprising a
dozen of OB stars and ∼ 30 OB star candidates around it
\citep{tsujimoto2007,rauw2011,zeidler2015}.  A binary Wolf-Rayet star
(WR20a), perhaps the most massive binary in the Galaxy, is associated
with the cluster and its presence indicates that the cluster is a few
10$^6$ years old. Beyond the cluster core are at least two very
massive stars, which include an O4 or O5 star and another Wolf-Rayet
star (WR20b).  More than 3000 X-Ray point sources were found centered
on Wd2 and toward its west lies a pulsar wind nebula surrounded by
diffuse X-Ray emission \citep{townsley2019}.  Stellar winds from Wd2
and from the surrounding stars play an important role in the star
formation occurring in RCW49. \\ The velocity dispersion of the
associated molecular gas suggests that collision between two CO
molecular clouds in the velocity ranges -11 to 9 km~s$^{-1}$ and 11 to
21 km~s$^{-1}$ also contributed to the formation of the stellar
cluster \citet{furukawa2009}. \citet{whiteoak1997} reported the
presence of two wind driven shells at the center of RCW49. We assume
that both the stellar winds and cloud-cloud collision are responsible
for the morphological evolution of the bubbles associated with
RCW49.\\

\noindent{\bf RCW79} \\
The RCW79 bubble is ionized by a cluster of a dozen O stars, the two
most massive of which have a spectral type O4-6V/III
\citep{martins2010}.  The ionizing luminosity of the ionizing stars
was estimated to be 10$^3$ times higher than the mechanical luminosity of
their stellar winds, indicating a radiation driven \HII\ region. RCW79
is spatially encompassed by an almost complete dust ring, with a
diameter of 12$'$, corresponding to $\sim$12 pc at a distance of 4.2
kpc \citep{russeil1998}. \citet{liu2017} suggested that triggered star
formation might occur around this bubble. Based on {\sl Herschel} data
for RCW79 \citep{liu2017}, more than 50 compact sources (Class 0 and
I) were found in the ionization-compressed layer of cold and dense gas
from which 12 are candidate massive dense cores that may form
high-mass stars.  The core formation efficiency (CFE) shows an
increase with increasing density, suggesting that the denser
the condensation, the higher the fraction of its mass transformation
into dense cores.  \\

\noindent{\bf RCW120} \\
This source is a well-studied, bubble-shaped \HII\ region of $\sim$4.5
pc diameter, excited by an O8V star, CD $-38^{\circ}11636$, at a
distance of 1.68~kpc \citep{kuhn2019}. RCW 120 has become sort of a
poster-child \citep{zavagno2010,anderson2015} of the myriad of bubbles
discovered by the GLIMPSE survey. The \HII\ region is surrounded by a
dense shell of gas and dust, observed in dust
\citep{zavagno2007,deharveng2009,anderson2015,figueira2017,zavagno2020}
and molecular lines \citep{torii2015,kirsanova2019}. Remarkable is an
arc of emission visible at 24~$\mu$m south of the ionizing star
because it may represent the upstream boundary between the wind bubble
and photoionized gas \citep{mackey2015}.  There is an ongoing
discussion whether the ring-shape appearance of RCW120 is due to an
expanding \HII\ region \citep{zavagno2007,deharveng2009} or a
cloud-cloud collision \citep{torii2015}.  Using {\sl Herschel}
observations \citet{anderson2012} found that 20\% of the total FIR
emission of bubble \HII\ regions comes from the direction of bubble
“interiors,” the locations inside the PDRs, which suggests a
three-dimensional morphology. \\

\noindent{\bf RCW36} \\
\noindent RCW36 is among the closest \HII\ regions to the Sun, at a
distance of 950 pc \citep{massi2019} within the Vela C molecular
cloud. The \HII\ region has a bipolar shape and is surrounded by
molecular gas with a dust lane that crosses the bipolar cavity. The
embedded cluster ($\sim$350 stars) with the most massive star being a
type O8 or O9 is located within the cavity \citep{baba2004}. The
cluster extends over a radius of 0.5 pc, with a stellar surface number
density of 3000 stars pc$^{−2}$ within the central 0.1 pc. {\sl
  Herschel} studies suggest that the bipolar morphology has evolved
from its filamentary beginnings under the impact of ionization
\citep{minier2013}.

The bipolar \HII\ region RCW36 is also an example of a region for the
interplay between ionization and structures (bright rims and pillars)
around an \HII\ region. Comparing the dynamics of \CII\ and CO
emission \citep{fissel2018} will address the question whether
filamentary structures can be the location of very dynamical phenomena
inducing the formation of dense clumps at the edge of
\HII\ regions. Moreover the \CII\ mapping of this region will lead to
a better understanding of the formation of bipolar nebulae as a
consequence of the expansion of an \HII\ region into a molecular ridge
or an interstellar filament. \\

\noindent{\bf W40} \\
The W40 complex is a nearby (260 pc, \citet{massi2019}) site of
high-mass star formation associated with a cold molecular cloud
($\sim$10$^4$~M$_{\odot}$) and includes a blistered \HII\ region
powered by an OB association and two interconnected cavities, forming
an hour-glass shape on large scales (a few pc). The main cluster is
located just northwest of the narrow waist where the two cavities are
joined. The bright-rimmed clouds at the cavity walls show clearly that
dense clumps and pillars are illuminated from inside by the cluster.
The OB association is comprised of IRS/OS1a (O9.5), IRS/OS2b (B4) and
IRS/OS3a (B3) and an associated stellar cluster of pre-main-sequence
stars \citep{shuping2012}.

The W40 molecular cloud/\HII\ region is one of the few nearby regions
with active high mass star formation, hosting an embedded cluster
\citep{koenyves2015}.  The \CII\ mapping will focus on the bipolar
cavity walls illuminated by the cluster to study the feedback of the
OB cluster on the surrounding molecular cloud.\\ \\

\noindent{\bf W43} \\
W43 is one of the most active star-forming regions in the Galaxy. Its
position in the Galactic plane and its radial velocity place it at the
junction point of the Galactic long bar and the Scutum spiral arm at
5.5 kpc distance to the Sun. It contains two of the most massive cloud
groups of the first Galactic quadrant (W43 Main and South) with a
total mass of $\sim$6$\times$10$^6$~M$_{\odot}$
\citep{carlhoff2013,motte2014}.  W43 Main is heated by a cluster of
Wolf-Rayet and OB stars ($\sim$3.5$\times$10$^6$~L$_{\odot}$) and is
considered to be a Galactic mini starburst region
\citep{motte2003,bally2010} since it is undergoing a remarkably
efficient episode of high-mass star-formation ($\sim$15 high-mass
protoclusters and a star-formation efficiency (SFE) of
$\sim$0.1~M$_{\odot}$ yr$^{-1}$).

The formation of the associated ionizing cluster was likely supported
by the expansion of an older \HII\ region to the south that triggered
the formation and gravitational collapse of the GMC that evolved into
W43. Subsequently, UV radiation from the central O+WR cluster
compressed the parent cloud toward both low- and high Galactic
longitudes, triggering the formation of additional massive stars. Its
mini starburst activity is continuously fueled by converging flows,
owing to its position at the junction point of the Scutum–Centaurus
(or Scutum-Crux) Galactic arm and the Bar
\citep{beuther2012,motte2014}.

Its position in the Galaxy makes W43 a very interesting object for
studying the formation of molecular clouds and the feedback in a
dynamically extreme environment. Despite its distance, it is possible
to analyze the details of this cloud, due to its large spatial scale
of 150 pc and the large amount of gas at high density.

\begin{figure*}
\centering
\includegraphics[width=14cm, angle=0]{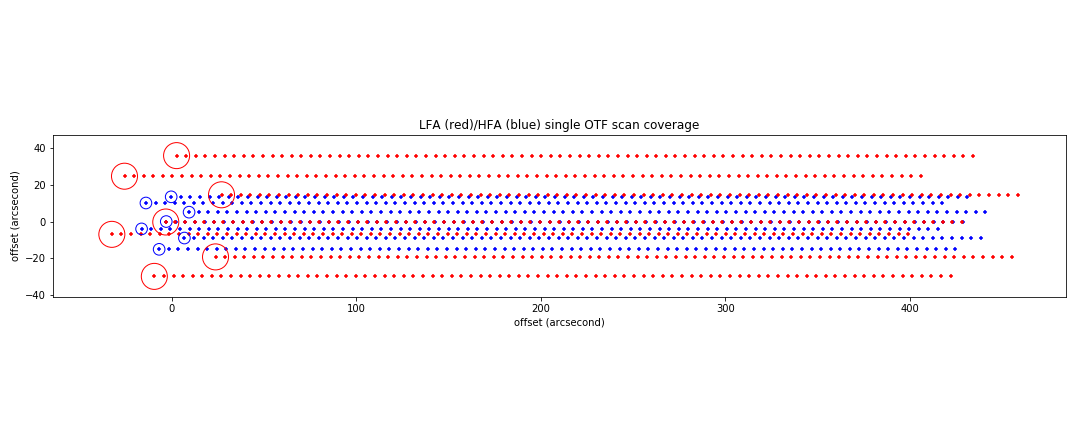}
\caption{LFA (red) and HFA (blue) footprints (diameter of the circles
  correspond to the FWHM of the beams) of the 7 pixel arrays (the LFA
  has 2$\times$7 pixels, one in horizontal and one in vertical
  polarization) as they appear on the sky for one scan.  The dotted
  lines show the scan direction, each dot represents one dump.  The
  array is rotated by 19$^\circ$ relative to the scan direction.  The
  LFA array size is 72.6$''$ (pixel spacing of 31.8$''$) and the HFA
  array size is 27.2$''$ (pixel spacing of 13.6$''$). }
\label{footprint}
\end{figure*}

\begin{figure*}
\centering \includegraphics[width=10cm, angle=-90]{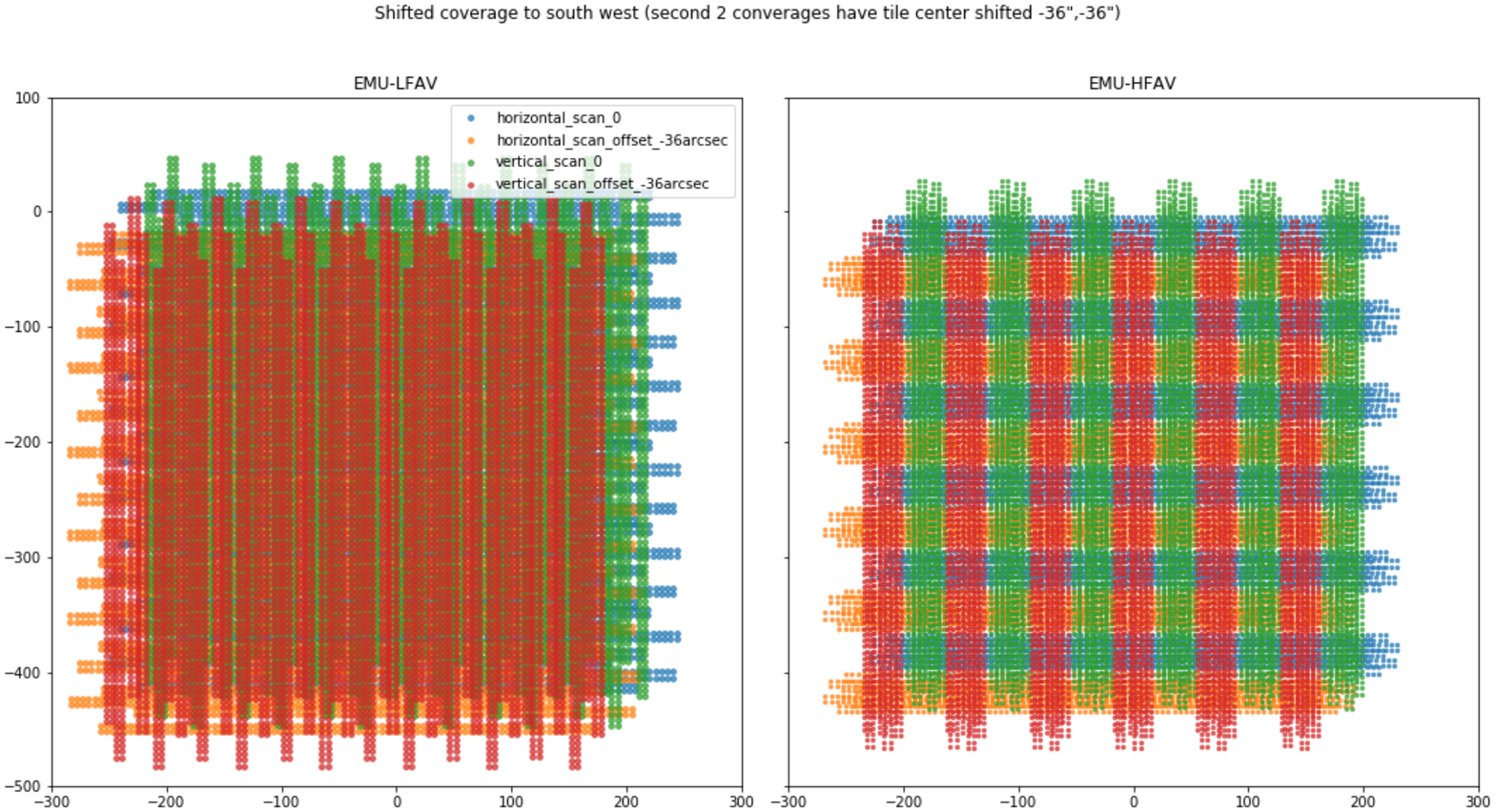}
\caption{Optimized array mapping for \CII\ and \OI\ observations. The
  left panel shows the 4 coverages for the LFA channel (\CII\ line)
  and the right panel the resulting coverage for the HFA channel
  (\OI\ line). The observations start with a horizontal scan (blue),
  followed by a vertical scan (green). Then, the tile center is
  shifted by -36$''$,-36$''$ and another horizontal and vertical scan
  (orange and red, respectively) are performed. The shift of the
  second coverage is chosen in order to fill the gap in the HFA map. }
\label{mapping}
\end{figure*}

\section{Observation strategy and planning} \label{observations}

\subsection{Definition of mapping area}

We have estimated the \CII\ 158 $\mu$m surface brightness from 8
$\mu$m PAH emission maps (convolved to the upGREAT beam) because the
latter are signposts of bright PDR gas and \CII\ and 8 $\mu$m
brightness are very well correlated with each other
\citep{pabst2017,pabst2019,anderson2019}.  The map extent was defined
by a certain expected level of \CII\ line integrated intensity
(varying between 100 and 250 K km s$^{-1}$, depending on source) and
compared to maps of the far-UV field, derived from 70 and 160 $\mu$m
Herschel/PACS flux maps (see procedure described in
\citet{schneider2016}). Typically, the mapping area starts above 100
G$_\circ$. Figure~\ref{sourcesfigure} shows the area to be covered by
the tiles of the \CII\ mapping overlaid on IRAC 8 $\mu$m maps in
color.  The emission free reference positions for each map are listed
in Table~\ref{table:summary1}.

\subsection{upGREAT} \label{upgreat} 
All sources are mapped with the dual-frequency heterodyne array
receiver upGREAT\footnote{German Receiver for Astronomy at
  Terahertz. (up)GREAT is a development by the MPI f\"ur
  Radioastronomie and the KOSMA/Universit\"at zu K\"oln, in
  cooperation with the DLR Institut f\"ur optische Sensorsysteme.}
heterodyne receiver \citep{risacher2018} with the \CII\ 158
\micron\ (1.9 THz) tuned in the LFA 2$\times$7 pixel array and the
\OI\ 63 \micron\ (4.7 THz) line tuned in the HFA 7 pixel array. The
LFA has an array size of 72.6$''$, and a pixel spacing of 31.8$''$
while the HFA has an array size of 27.2$''$ and a pixel spacing of
13.6$''$ (see Figure~\ref{footprint} for a footprint of the arrays on
the sky during one scan).  The half-power beam widths are 14.1$''$
(1.9 THz) and 6.3$''$ (4.7 THz), determined by the instrument and
telescope optics, and confirmed by observations of planets. The
receiver noise temperatures are 2000~K for the LFA and 2500~K for the
HFA (see Table 1 in \citet{risacher2018}).
%
%
%
The backends for both channels are Fast Fourier Transform Spectrometer
(FFTS) with 4 GHz instantaneous bandwidth \citep{klein2012}. The
frequency resolution of the raw  \CII\ and \OI\ data is hardware selected to 0.244 MHz,
giving a velocity resolution of 0.04 km s$^{-1}$ and 0.015 km s$^{-1}$, respectively.
The data cubes provided to the SOFIA science center and
used by the FEEDBACK consortium have a resampled velocity resolution
of 0.2 km s$^{-1}$.


Procedures to determine the instrument alignment and telescope
efficiencies, antenna temperature and atmospheric transmission calibration, as well
as the spectrometers used, are described in \citet{risacher2016,risacher2018} and
\citet{guan2012}. For each flight series, the main-beam efficiencies
($\eta_{mb}$) for each pixel for the LFA and HFA channels are
determined, typical values are $\eta_{mb}$=0.65 (0.69) for \CII\ (\OI).
The forward efficiency is $\eta_{f}$=0.97. 




The telluric \OI\ line, originating from the mesosphere, can
contribute as a narrow feature in the observed band. The half-intensity width
of this non-gaussian line is typically 1 km s$^{-1}$. We try to
schedule each source at an observing time of the year when the mesospheric line
does not appear at the velocity of the bulk emission.  However,
because of flight planning constraints and because the \CII\ line is 
the main science driver, this is not always possible.
We will endeavor to maximize the scientific return around the
atmospheric line, but the line can be rather opaque at the line center.
We will need to fine tune the atmospheric model layering of the mesophere to determine
the correct transmission on the wings of the atmospheric \OI\ line  
profile. Atomic oxygen is now included in the latest
version of the 'am' atmospheric code \citep{paine2019} used to calibrate the upGREAT 
data. The large FEEDBACK data set should give us 
ample data to constrain the distribution of atomic oxygen in the 
atmosphere (H\"ubers et al. 2020, sub.). \\
We will also consider to apply a correction for the absorption following 
the procedure described in \citet{leurini2015} and 
\citet{schneider2018}, assuming that the profile can be characterized 
by a Gaussian. In summary, for the opacity correction, the absorption 
strength was adjusted in a way to achieve an adequate interpolation 
between adjacent unaffected spectral channels. \\
The thus corrected \OI\ data will be delivered later as a Level 4 data product. 

\subsection{Mapping scheme} \label{scheme}
Mapping was performed in an optimized array-on-the-fly mapping
mode. Each region was split into multiple square ’tiles’ with
435.6$''$ on one side and each square was covered 4 times.  The OTF
scan speed was selected to attain Nyquist sampling of the LFA beam
(dump every 5.2$''$) while the HFA is undersampled.  The total time
for one OTF line is then 25.2s. This is, together with the OFF
observation, within the measured Allan variance stability time of the
system. The Allan variance of the LFA system is of the order of
80--100s under stable ambient temperature conditions but can decrease
to 30-35s if there are temperature instabilities or drifts onboard the
SOFIA aircraft \citep{risacher2018}.  The first two coverages are done
once horizontally and vertically with the array rotated 19$^{\circ}$
against the scan direction, so that scans by 7 pixels are equally
spaced.  The second two coverages are then shifted by 36$''$ in both
directions to achieve the best possible coverage for the \OI\ line in
the LFA array mapping mode (Figure \ref{mapping}).  Each tile takes
$\approx$50 minutes to complete.  This mode of operation is a slightly
modified version of the mode used in the \CII\ survey of the Orion
molecular cloud \citep{pabst2019} and Higgins et al., in preparation,
where the second coverage was performed with rotating the hexagonal
array by 60$^\circ$ instead of shifting the mapping positions. The
mode used in the FEEDBACK project achieves a higher degree of
redundancy as each pixel maps a different strip of sky during each
scan. Most importantly, though, this mode returns a better, though not
fully sampled, coverage of \OI.  The only disadvantage is that the
\CII\ maps may show edge effects since some LFA coverage is sacrificed
toward 2 sides of the map edges. However, there is still a uniform
noise coverage for the LFA.  At the beginning of each flight leg for
the project, a short calibration observation of a single point in each
source is performed to monitor the line intensity if the observations
extend over several flights.

\subsection{Data reduction} \label{data}

Spectra are presented on a main beam brightness temperature scale
T$_{mb}$. Recommended main beam coupling efficiencies are presented in
the notes "Project Overview" in the SOFIA science archive.  All pixels
have been calibrated individually and there are slight efficiency
variations depending on the Cycle when the sources were observed. For
the sources observed in Cycle 7 and 8, the main-beam effciencies are
typically 0.65 for \CII\ and 0.69 for \OI.  The calibrated \CII\ and
\OI\ spectra were further reduced and analyzed with the
GILDAS\footnote{GILDAS is developed and maintained by IRAM.}
software. From the spectra, a 3rd order baseline was removed and the
spectra were then gridded with 1/$\sigma^2$ weighting (average
baseline noise).

\begin{figure}
\centering \includegraphics[width=8cm, angle=0]{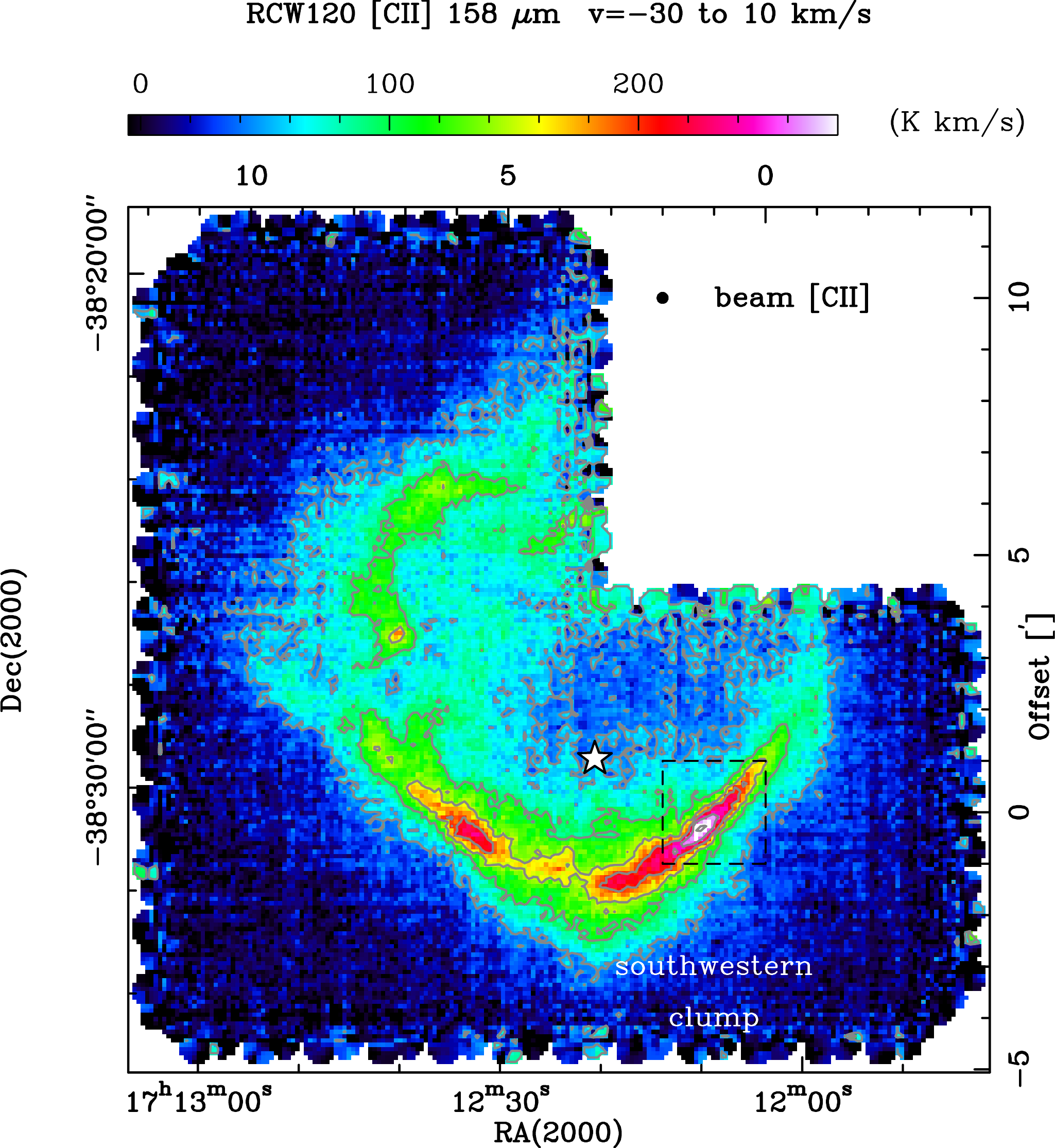}
\caption{SOFIA/upGREAT map of line integrated \CII\ emission in RCW120 smoothed to an
  angular resolution of 15$''$ and with a gridding of 3.5$''$. The star marks
  the position of the exciting O8V star CD38-11636. The
  dashed rectangles indicate the area used for averaging the 
  spectra presented in Fig.~\ref{rcw120-spectra}.}
\label{rcw120-int}
\end{figure}

\begin{figure}
\centering \includegraphics[width=8cm, angle=0]{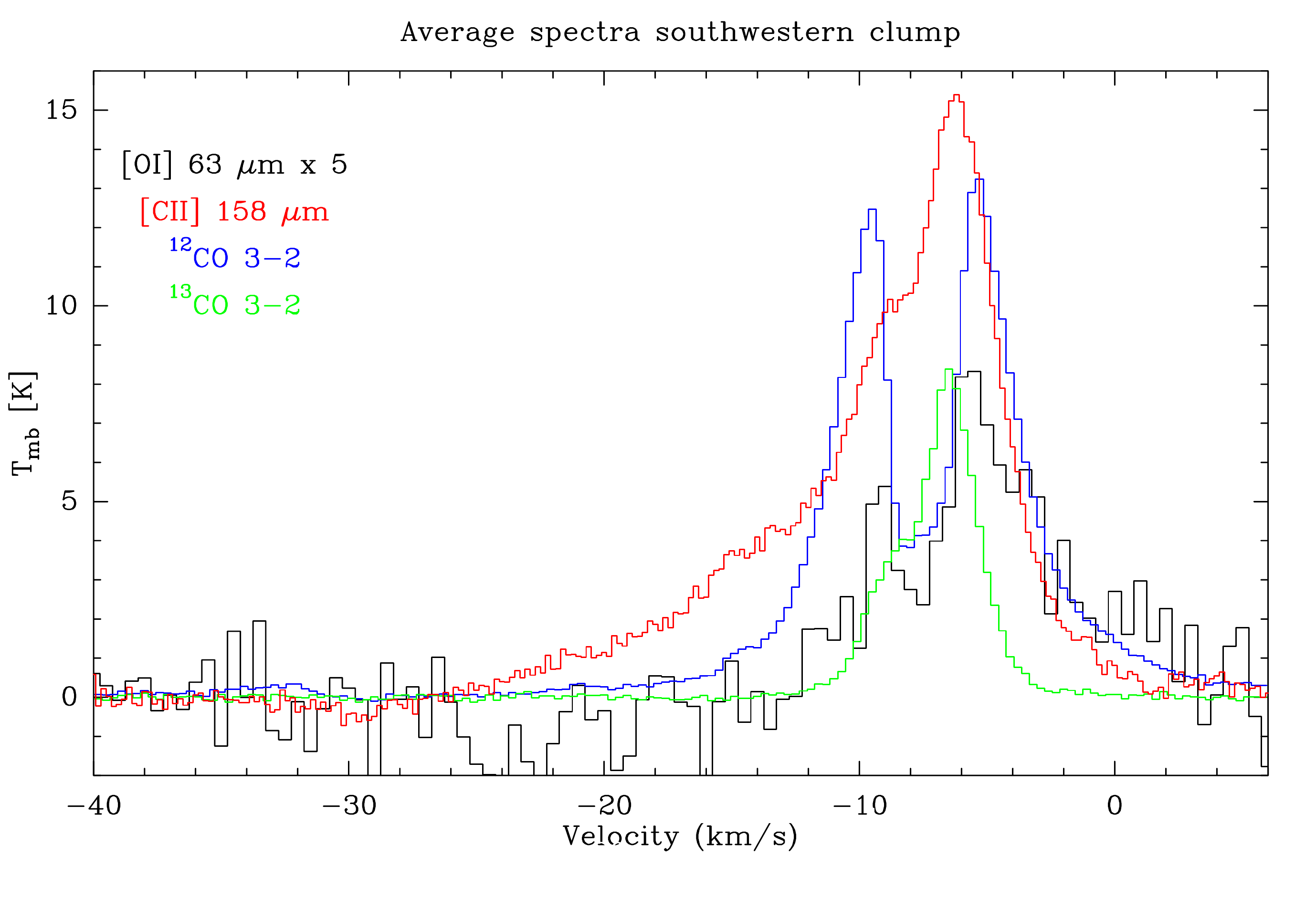}
\caption{Positionally averaged spectra of \CII\ (red), \OI\ (black),
  $^{12}$CO 3$\to$2 (blue), and $^{13}$CO 3$\to$2
  (green). Figure~\ref{rcw120-int} shows the area of integration in
  the southeastern clump.  Note that the \OI\ line temperature is
  multiplied by 5 for better visibility. The \OI\ and $^{12}$CO lines show a strong dip
  at -8~km s$^{-1}$ which is also visible - though less prominent - for the \CII\ and $^{13}$CO 3$\to$2 lines.}
\label{rcw120-spectra}
\end{figure}

\subsection{RCW120 data} \label{datarcw120}

The example data we show in this paper (Figs.~\ref{rcw120-int} and
\ref{rcw120-spectra}) were obtained during one flight from
Christchurch, New Zealand on 10th of June 2019 (flight F579). Three
out of the four planned tiles were observed in the mapping scheme
described in Sec. \ref{scheme} (the last tile was not completed so
that 72\% of the map is finished). The data were reduced as outlined
in Section \ref{data}. The \OI\ mesospheric line at $\sim$8 km
s$^{-1}$ was not much affecting the astronomical \OI\ line at around
-8 km s$^{-1}$. We here show preliminary data that were obtained by
removing a third order baseline, and excluding the velocity range of
the \OI\ telluric line.

\section{First results} \label{results}

\subsection{Current state of observations } \label{current}

As of April 2020, observations for 32\% of the 96h of the observing time of
the FEEDBACK program have been performed.  The state of observations
is constantly updated and is shown on the Cologne FEEDBACK webpage.
The most advanced map is that of RCW120 (Section \ref{rcw120}) with
72\% completed. The \HII\ bubble RCW49 is 65\% complete, and RCW36
in Vela 51\%. For all other sources, the completeness is less than 20\%.

\subsection{\CII\ and \OI\ observations of RCW120} \label{rcw120}

We selected the RCW120 region as an example for the mapping
capabilities of upGREAT on SOFIA because these observations are the
most progressed. The maps and spectra shown in this subsection are
preliminary and intended to give a first impression of the nature and
the quality of the data that the FEEDBACK program will deliver. It is
not our objective here to discuss in more detail the science, which
will be done in Luisi et al. (in preparation) and other upcoming
publications.  Figure~\ref{rcw120-int} shows the line integrated
\CII\ map, smoothed to 15$''$ angular resolution (the \CII\ beam is
14.1$''$) and on a 3.5$''$ grid.  The velocity range for integration
was -30 to +10 km s$^{-1}$ which covers all relevant velocity ranges
of \CII\ emission. The overall structure is shell-like with two
emission peaks in the southwest (up to $\sim$300 K km s$^{-1}$) and
southeast (up to $\sim$200 K km s$^{-1}$), and more diffuse emission
from the bubble center. These two peaks correspond to condensations 1
and 2 labeled by \citet{zavagno2007}. \\ Figure~\ref{rcw120-spectra}
displays an averaged \CII\ spectrum over the southwestern clump,
indicated in Fig.~\ref{rcw120-int} as a dashed polygon, together with
spectra of \OI\ 63 $\mu$m, and $^{12}$CO 3$\to$2 and $^{13}$CO 3$\to$2
data from APEX, averaged over the same area. Note that the \OI\ line
shows some baseline structures because we conservatively only removed
a baseline order 3 for this preliminary data reduction. The
\CII\ spectrum shows a prominent blue tail, reaching up to -30 km
s$^{-1}$ which is not apparent in the other lines.  All spectra have a
double-peak structure which is most prominent for the $^{12}$CO
3$\to$2 and the \OI\ line with a deep dip at $\sim$-8 km
s$^{-1}$. Because the \CII\ line and the $^{13}$CO 3$\to$2 line (which
is commonly considered to be optically thin) also show a dip at that
velocity - though not as deep as the one for \OI\ and $^{12}$CO - the
interpretation of this profile can be twofold. Either all 4 lines are
affected by self-absorption, or there are individual velocity
components. Self-absorption is the most likely explanation, though it
requires a complex layering of various gas components at different
densities and temperatures along the line of sight because the
excitation conditions for all lines are distinct
(Sec.~\ref{introduction}). While the \CII\ line is thermally excited
at temperatures around 100~K and densities of a few 10$^3$ cm$^{-3}$,
the CO lines require higher densities (a few 10$^4$ cm$^{-3}$), and
the \OI\ line has the highest critical density of 10$^5$ cm$^{-3}$ and
necessitates temperatures $>$200~K. A temperature gradient clearly
exists, with higher values in the PDR facing the inside of the
shell. {\sl Herschel} dust observations \citep{anderson2010} show
temperatures $>$30~K in the interior of RCW120 (with a hot dust
component around 100 K), $\sim$20~K in the PDR, and $\sim$10~K in the
cool molecular shell.  In addition, YSOs in different evolutionary
phases from Class 0 to Class I were detected in the southwestern clump
\citep{figueira2017} and could contribute as internal heating
sources. Clumpiness of the shell would lead to higher density, cool
clumps embedded in a lower density, warm interclump
medium. Self-absorption in the observed CO lines requires cool,
low-density gas. \citet{kirsanova2019} fitted their $^{13}$CO 3$\to$2
and 2$\to$1 observations with a foreground cloud with a density of
about 50 cm$^{−3}$ and a temperature $<$60 K. This gas component would
also lead to self-absorption in the \CII\ line.  A more detailed
investigation is out of the scope of this paper, but will be done in a
furthcoming study (Kabanovic et al., in preparation). We will then also use
observations of the \tCII\ line to determine the optical depth of the
\twCII\ line and employ models of several emission and absorption
layers, similar as it was done in \citet{guevara2020} for M17.

\section{Data products} \label{products}

For each source, Level 3 data products with calibrated \CII\ maps are
delivered to the SOFIA Science Center by the GREAT team.
The data are accessible via the NASA/IPAC InfraRed Science
Archive\footnote{\href{https://irsa.ipac.caltech.edu/applications/sofia}{https://irsa.ipac.caltech.edu/applications/sofia}.}
No account is required to access the archive, and the FEEDBACK data are
available to the public upon their ingestion. FEEDBACK data can be
accessed at IRSA by searching by source (under 'Spatial Constraints')
or by Plan ID 07\_007 (under 'Proposal Constraints'). The Level 3  
data in the archive were produced by removing a third order
spectral baseline, calibration to T$_{mb}$ temperature scale,
resampling to 0.2 km s$^{-1}$ spectral resolution, and filtering using
the ratio of (baseline rms)/(expected radiometer noise) $<$1.40.
Level 4 data, for which further issues such as, e.g., standing waves or the
mesopheric \OI\ line are addressed, will be delivered later.

\section{Conclusions} \label{conclusions}

We presented the goals and first promising results of the SOFIA legacy
project FEEDBACK that is currently mapping the \CII\ 158 $\mu$m and
the \OI\ 63 $\mu$m fine structure lines in 11 Galactic star-forming
regions with the upGREAT instrument.  The \CII\ line uniquely provides
the kinematics of the gas exposed to the mechanical and radiative energy input by
massive stars.  By surveying regions with a range of massive star
formation activity with stars of different spectral type, we will
quantify the relationship between star formation activity and energy
injection. We assess the negative (inhibition of star-formation) and
positive (triggering of star-formation) feedback processes involved,
and link that to other measures of activity on scales impacted by individual
massive stars, small stellar groups, and star clusters.
The \OI\ line serves as a tracer for high density, high temperature PDRs and
for shocks. 

FEEDBACK takes full advantage of the unique capabilities of the
upGREAT/SOFIA combination: The high spatial (14.1$''$) and spectral
(0.2 km s$^{-1}$) resolution of the 14 pixel LFA upGREAT heterodyne
spectrometer coupled with the nimble telescope of SOFIA allows for
efficient mapping of the \CII\ line over large (100's to 1000's of
square arcmin) areas. With a total observing time of 96h, we will
cover $\sim$6700 arcmin$^2$ in the selected sources which are Cygnus
X, M16, M17, NGC6334, NGC7538, RCW49, RCW79, RCW120, RCW36, W40, and W43.
These \CII\ maps, together with data for the less explored \OI\ line,
are delivered by the FEEDBACK consortium and are publically
available. They provide a comprehensive database for the astronomical
community and will serve as a starting point for many studies and
follow-up observations.

\section{Acknowledgments}
This work is based on observations made with the NASA/DLR
Stratospheric Observatory for Infrared Astronomy (SOFIA). SOFIA is
jointly operated by the Universities Space Research Association,
Inc. (USRA), under NASA contract NNA17BF53C, and the Deutsches SOFIA
Institut (DSI) under DLR contract 50 OK 0901 to the University of
Stuttgart. Financial support for the SOFIA Legacy Program, FEEDBACK,
at the University of Maryland was provided by NASA through award
SOF070077 issued by USRA. \\ 
We thank the USRA and NASA staff of the Armstrong Flight Research
Center in Palmdale and of the Ames Research Center in Mountain View,
and the Deutsches SOFIA Institut for their work on the observatory. We
acknowledge the support by the upGREAT team for operating the
instrument, for planning the detailed observing scenarios and in the
calibration of the data. The development and operation of upGREAT was
financed by resources from the MPI f\"ur Radioastronomie, Bonn, from
Universit\"at zu K\"oln, from the DLR Institut für Optische Sensorsysteme,
Berlin, and by the Deutsche Forschungsgemeinschaft (DFG) within the
grant for the Collaborative Research Center 956 as well as by the Federal Ministry
of Economics and Energy (BMWI) via the German Space Agency (DLR)
under Grants 50 OK 1102, 50 OK 1103 and 50 OK 1104. \\ 
The FEEDBACK project is supported by the BMWI via DLR, Projekt Number 50 OR 1916
(FEEDBACK) and Projekt Number 50 OR 1714 (MOBS - MOdellierung von
Beobachtungsdaten SOFIA).\\
This work was also supported by the Agence National de Recherche
(ANR/France) and the Deutsche Forschungsgemeinschaft (DFG/Germany)
through the project "GENESIS" (ANR-16-CE92-0035-01/DFG1591/2-1). \\ 
Interstellar medium studies at Leiden Observatory
are supported through a Spinoza award. \\
H.B. acknowledges support from the European
Research Council under the Horizon 2020 Framework Program via the ERC
Consolidator Grant CSF-648505. H.B. also acknowledges support from the
DFG via SFB 881 “The Milky Way System” (sub-project B1).

\bibliographystyle{aa}
\newcommand{\newblock}{}

\end{document}